\documentclass[article,longmktitle,authoryear]{cas-sc}

\usepackage[T1]{fontenc}
\usepackage[utf8]{inputenc}
\usepackage[english]{babel}

\usepackage[numbers]{natbib}
\usepackage{xpatch}
\usepackage{dsfont}
\usepackage{mathrsfs}
\usepackage{url}            
\usepackage{booktabs}       
\usepackage{amsfonts}       
\usepackage{nicefrac}       
\usepackage{microtype}      
\usepackage{lipsum}         %
\usepackage{caption}
\usepackage{graphicx}       %
\usepackage{color}          %
\usepackage{amsmath}
\usepackage{amssymb}
\usepackage{caption}
\usepackage{multirow}
\usepackage{multicol}
\usepackage{times}
\usepackage{graphicx}
\usepackage{breakcites}
\usepackage{natbib,enumitem}
\usepackage{bm} 
\usepackage{bbm}
\usepackage{epsfig}
\usepackage{alltt}
\usepackage{tikz}
\usepackage{booktabs}
\usepackage{pgfplots}
\pgfplotsset{compat=1.6}
\usepackage{xcolor}

\usepackage{algorithm}
\usepackage{algorithmic}
\usepackage[ruled,vlined, algo2e]{algorithm2e}
\usepackage{tcolorbox}
\usepackage{tikz}
\usepackage{tikz}
\usepackage{tikz-qtree,tikz-qtree-compat}

\newtheorem{defn}[subsection]{Definition}

\def\tsc#1{\csdef{#1}{\textsc{\lowercase{#1}}\xspace}}
\tsc{WGM}
\tsc{QE}
\tsc{EP}
\tsc{PMS}
\tsc{BEC}
\tsc{DE}

\begin{document}
\let\WriteBookmarks\relax
\def\floatpagepagefraction{1}
\def\textpagefraction{.001}
\shorttitle{}
\shortauthors{Cyrille Feudjio et~al}

\title [mode = title]{A Novel Use of Discrete Wavelet Transform Features in the Prediction of Epileptic Seizures 
from EEG Data}                      
	\tnotemark[1]
	
	\tnotetext[1]{This document is the results of the research
	 project funded by AIMS CAMEROON with the help of Mastercard Foundation.}
	

\author[1]{Cyrille Feudjio}[
	orcid=0000-0003-4627-9214]
\cormark[1]
\ead{cyrille.feudjio@aims-cameroon.org}
	
\credit{Investigation, Conceptualization, Methodology, Software, Data curation, Experimentation, Result compilation, Writing - original draft}
	
\address[1]{School of Mathematical Sciences, African Institute for Mathematical Sciences, Crystal Gardens, Limbe Cameroon}
\address[2]{School of Mathematical Sciences, Stellenbosch University, South Africa}
\address[3]{School of Mathematical Sciences, Rochester Institute of Technology, Rochester, NY 14623}
	
\author[1]{Victoire Djimna Noyum}
\ead{victoire.djimna@aims-cameroon.org}
	
\credit{Software, Result compilation, Writing - review}
	
\author[1]{Younous Perieukeu Mofendjou}
\ead{younous.mofenjou@aims-cameroon.org}
	
\credit{Software, Result compilation, Writing - review }
	
\author[2]{Rockefeller}
\ead{rockefeller@aims-senegal.org}
	
\credit{Supervision, Validation, Writing - review \& editing}

\author[3]{Ernest Fokou\'e}
\cormark[1]
\ead{epfeqa@rit.edu}
	
\credit{Conceptualization of this study, Supervision, Validation, Writing - review \& editing}

\cortext[cor1]{Corresponding author}
	
\nonumnote{In this work, we demonstrate the predictive superiority of discrete wavelet transform over previously used methods of feature extraction in the diagnosis epileptic seizures from EEG data.
	}
	
\begin{abstract}
	This paper demonstrates the predictive superiority of discrete wavelet transform (DWT) over previously used methods of feature extraction in the diagnosis of epileptic seizures from EEG data. Classification accuracy, specificity, and sensitivity are used as evaluation metrics. We specifically show the immense potential of 2 combinations (DWT-db4 combined with SVM and DWT-db2 combined with RF) as compared to others when it comes to diagnosing epileptic seizures either in the balanced or the imbalanced dataset. The results also highlight that MFCC performs less than all the DWT used in this study and that, The mean-differences are statistically significant respectively in the imbalanced and balanced dataset. Finally, either in the balanced or the imbalanced dataset, the feature extraction techniques, the models, and the interaction between them have a statistically significant effect on the classification accuracy.
	
\end{abstract}
	
	
	

\maketitle
\section{Introduction}~
Nowadays, people face various kinds of stress in their daily lives and most of the people around the world suffer from a range of neurological disorders. Epilepsy affects up to 1\% of the population, making it, alarmingly, the third commonest encephalopathy \citep{usman2017epileptic}. It can affect both males and females of all races, ethnic, backgrounds, and ages.

Approximately 50 million people worldwide suffer from epilepsy and 90\% of them are from developing countries\citep{kandar2012epilepsy}. It is not one disorder but rather a syndrome with widely divergent symptoms involving episodic abnormal electrical activity within the brain. Patients with epilepsy could be treated with medication or surgical procedures \citep{guenot2004surgical}. However, these methods are not fully effective. Unfortunately, seizures that can't be fully treated medically limit the patient's active life, and in these cases, patients cannot work independently and perform certain activities. This ends up in the social isolation of people and economic difficulties. However, early prediction of epileptic seizures can ensure sufficient time before they occur. 

Tons of effort has been put in situ by researchers and institutions to modify that.But interestingly, the main cause of it remains a mystery. Only an early diagnosis pauses as a secure and plausible way to treat it. Therefore, several methods are developed to detect an epileptic seizure before it starts. Machine learning models are used for this task which incorporates Electroencephalography (EEG) signal acquisition and preprocessing, features extraction from the signals, and finally, classification between different seizure states. Electroencephalography (EEG) may be a useful method to watch the nonlinear electrical function of the brain's nerve cells; hence, it is a valuable tool for the epilepsy evaluation and treatment\citep{wang2013comparison}.

Feature extraction which involves tidying the data is usually said to represent where 80\% of the time is spent working on a data science. In this case, for instance, Preprocessing and feature extraction from EEG signals have an excellent effect on maximizing prediction time. Literature~\citep{rasekhi2013preprocessing},\citep{teixeira2014epileptic},\citep{bandarabadi2015epileptic},\citep{zandi2013predicting} states that no machine learning model provides a reliable method for pre-processing and feature extraction. However, each of these processes is tailored to specific problems, which makes them indispensable before building the model. Therefore, this project aims to look at the predictive effect of feature extraction methods within the EEG dataset especially in the case of an epileptic seizure.

\subsection{Review on Epilepsy Seizure Classification Using EEG data}~

Some researchers conducted studies to detect the phase of seizures through EEG signal processing as reported within the study of \citep{ullah2018automated}. During this study, the classification of the seizure phase was called pre-ictal, ictal, inter-ictal, and postictal to analyze differences in characteristics in each phase. The tactic used for signal processing was done in the time, frequency or time-frequency domains. Furthermore, another important study was early detection of the phase before seizures to supply an alarm to epilepsy patients as reported within the study of ~\citep{saputro2019seizure}. During this research, they detected the kind of seizures as opportunities and challenges to help the neurologist in classifying the seizure from EEG recording.

\citep{golmohammadi2017tuh} conducted an epileptic EEG signal processing simulation to differentiate the types of seizures. The seizure types studied during this research were a generalized non-specific seizure, non-specific seizure, and tonic-clonic seizure. The methods utilized during this research were Mel Frequency Cepstral Coefficients (MFCC), Hjorth Descriptor, and Independent Component Analysis (ICA) as features extractions while Support Vector Machine (SVM) was used as a classifier.

~\citep{mursalin2017automated} presented a hybrid approach where features from time and frequency domains were analyzed to detect epileptic seizures from EEG signals. They started by applying an Improved Correlation-based Feature Selection (ICFS) method to capture relevant features from the time domain, frequency domain, and entropy-based features. Then, the classification of the selected feature was done by an ensemble of Random Forest (RF) classifiers. Results revealed that the proposed method was better in performance as compared to the conventional correlation-based and some other state-of-the-art methods of epileptic seizure detection.

An automatic epilepsy diagnosis framework based on the combination of multi-domain feature extraction and nonlinear analysis of EEG signals was proposed by ~\citep{wang2017automatic}. EEG signals were pre-processed by using the wavelet threshold method to remove the artifacts, and representative features within the time domain, frequency domain, time-frequency domain, and nonlinear analysis features was extracted. The optimal combination of the extracted features was identified and evaluated via different classifiers. Experimental results demonstrated that, the proposed epileptic seizure detection method can achieve a high average accuracy of 99.25\%.

keeping in mind the fact that preprocessing of the EEG signals can improve prediction sensitivity and average anticipation time,~\citep{saputro2019seizure} proposed an efficient machine learning method for epilepsy prediction. In their research, they classified three sorts of seizure; Generalized Non-Specific Seizure (GNSZ), Focal Non-Specific Seizure (FNSZ), and Tonic-Clonic Seizure (TCSZ). They used the mixture of three feature extraction methods, Mel Frequency Cepstral Coefficients (MFCC), Hjorth Descriptor and Independent Component Analysis (ICA). The most effective result was obtained by combining MFCC and Hjorth descriptors which detected seizure type with 91.4\% as average accuracy.

~\citep{paul2018various} proposed a method for automatic seizure detection based on the mean and minimum value of energy. The algorithm was tested on the CHB-MIT database on three subjects with 60 and 40\% of data used as training and test data, respectively. They obtained an average detection accuracy of 99.81\%

\citep{bandarabadi2015epileptic} proposed an algorithm to predict epilepsy seizures which can extend the lifetime of epilepsy-affected patients. they extracted spectral power features, and after an appropriate selection, these features were passed into Support Vector Machines for classification. They observed sensitivity of 75.8\% and concluded that, reducing the proposed features subset can improve seizure prediction performance.

\citep{teixeira2014epileptic} proposed a model for the prediction of epileptic seizures by choosing  six channels of EEG signals and extracted 22 linear univariate features for each channel. They tested their model for prediction by varying multiple combinations of electrodes and also with four different pre-ictal state durations. They used three classifiers and approximately predicted every seizure. After selecting suitable features, training data was fed into Support Vector Machine for training, then test data was passed for determining classification accuracy and sensitivity. They observed 75.8\% as sensitivity of detecting the seizure.

Many researchers used EEG signals to detect the beginning of the pre-ictal state of epilepsy. However, only a few have reliably detected it .\citep{rasekhi2013preprocessing} proposed an algorithm for seizure prediction with the help of univariate linear features. They used six EEG channels in their proposed model and extracted 22 univariate linear properties. Support Vector Machine was used as a classifier to classify the preictal and ictal states of EEG signals. On the average , the prediction sensitivity after applying this algorithm was 73.90\%.

\citep{gadhoumi2012discriminating} used a wavelet method for the prediction of seizures. They extracted features including wavelet energy and wavelet entropy. Two or three channels were selected for testing purposes on a dataset of six patients. Sensitivity was reported as 88\% with a mean anticipation time of twenty-two minutes.

\citep{bhople2012fast} proposed an epileptic seizure detection method by using a Fast Fourier Transform (FFT). The FFT-based features were extracted and were fed to the neural networks. Multilayer perceptron (MLP) and Generalized Feed-Forward Neural Network (GFFNN) were used as a classifier. The algorithm was tested on the Bonn database, and results show they were able to achieve 100\% accuracy.

\citep{acharya2012application} designed a method for the detection of three states of EEG signal, (normal, pre-ictal, and ictal conditions) from recorded EEG signals. They combined the features from two domains; time domain, and frequency domain, and found that this combined features method is performing well in situations where the signal has a nonlinear and non-stationary nature.

\subsection{Contribution}~

Early detection is an important step in assisting people with epilepsy to take preventive measures against the upcoming manifestation of the disease/disorder, such as finding a secure place before the seizures occur. Classification of seizure could be a significant milestone in the journey through a potential or proper treatment, and if possible, prognosis prediction. In this regard, several automatic methods for detecting epileptic activity have been proposed recently ~\citep{wang2017automatic}. Most of them use Fourier Spectral Analysis for the extraction of EEG signals under the assumption that EEG signals are stationary~\citep{polat2007classification}, allowing the transformation of signals from the time domain to the frequency domain. Also, wavelet transformation approaches for time-frequency estimation are generally interesting. For instance, the Discrete Wavelet Transform (DWT) method which is a classical method of time-frequency analysis similar to the Short-Term Fourier Transform has been used to extract features from EEG signals~\citep{acharya2012application}. In addition to the extraction of time-frequency characteristics, non-linear analysis of EEG signals have also received considerable attention for detecting seizures which can be considered as a transition of the human brain~\citep{gajic2015detection}. We also have several discrete wavelet transformations using multi-domain characteristics and non-linear analysis to improve the performance of EEG seizure detection. Besides that, other feature extraction methods such as Mel frequency cepstral coefficients (MFCC), Hjorth descriptor, and Independent Component Analysis (ICA) could also be genuinely used. All of these methods are designed to remove redundant and irrelevant features so that the classification of new instances becomes more accurate. Researchers continue to explore these methods because the accuracy or sensitivity of classification models is highly dependent on the features used for prediction. Therefore, our contribution is to establish or at least intelligently speculate on the predictive powers of the feature extraction methods used. Tackling this will help us to:
\begin{itemize}
	\item Learn about different Machine Learning methods that can be combined with the feature extraction process and interpret the outcomes to build a kind of hybrid model that could hopefully generalize well.
	\item Potentially build a whole method that works well on the EEG dataset and could be extended to other domains where time series or wave signals are used.
	\item In the long run, build a package to make the whole process (feature extractions, fitting, and evaluating models) easy to use through dialogue boxes; both for medical purposes and social good.
\end{itemize} 
Contextually, the focus throughout the study will be on MFCC and the 3 best Wavelets as feature extraction methods.

\section{Feature Extraction Techniques}~
Feature extraction techniques are methods that select and/or combine variables into relevant features, effectively reducing the quantity of information that has got to be processed, while still accurately and completely describing the original dataset. In this chapter, we present two features extraction techniques used on the EEG dataset namely Wavelet Transform and Mel Frequency Cepstral Coefficient(MFCC).

\subsection{Wavelet Transforms}~

\begin{figure}[h]
	\centering
	\captionsetup{justification=centering,margin=2cm}
	\includegraphics[width=0.20\textwidth]{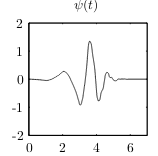}~\includegraphics[width=0.20\textwidth]{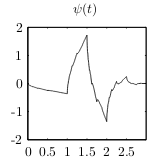}~\includegraphics[width=0.20\textwidth]{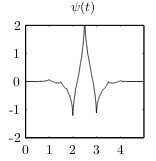}
	\caption{Kind of wavelet functions}
	\label{fig:f_wave}
\end{figure}

A wavelet $\psi(t)$ is a small wave, which must be oscillatory in some way to discriminate between different frequencies~\citep{merry2005wavelet}. It allows complex information content to be decomposed into elementary form at different positions and scales, and subsequently reconstructed back again with high accuracy. Figure~\ref{fig:f_wave} shows some examples of a possible wavelet.

The use of Wavelet Transform can work continuously (CWT) or discrete (DWT).

Given a time-domain signal function $f(t)$, and a wavelet function $\psi(t)$, the Continuous Wavelet Transform is defined by equation~(\ref{eq31}).
\begin{eqnarray}\label{eq31}
\Psi_f^{\psi}(\tau,s)=\dfrac{1}{\sqrt{|s|}}\int_{-\infty}^{+\infty} f(t)\psi^{*}\left(\dfrac{t-\tau}{s}\right)dt \ ,
\end{eqnarray}

Where $\tau$ and $s$ represent the translation and scale parameters respectively while $\psi(t)$ is called the mother wavelet. The symbol * indicates that, in case of a complex wavelet, the complex conjugate is used. By discretizing these parameters, the DWT is obtained~\citep{al2020general}.

Several Wavelets Transform namely Discrete Wavelet Transforms (DWT),  Discrete Fourier Transform (DFT), Single Valued Decomposition (SVD), Empirical Mode Decomposition (EMD), and their variants are widely used for seizure detection and prediction applications. Although many other time-frequency feature engineering approaches are prevailing for signal processing such as EMD, SVD, ICA, and PCA~\citep{al2020general}, DWT based wavelet feature analysis is identified as effective for time-frequency domain analysis because of its multiscale approximation feature. Another highlight of DWT based feature engineering is that, it is employed for both signal noise reduction as pre-processing, and feature extraction. The main characteristic of DWT which makes it the most effective method for analysis of EEG signals is its resolution of frequency and time. This property leads to optimality status for frequency-time resolution~\citep{chen2017high}. However, there exist many families of DWT transform described below.

\subsubsection{Family of wavelets}~

There are many types of DWT, which are considered as mathematical and statistical functions. These types are divided into families according to frequency components. Seven different types of common wavelets appearing in the literature~\citep{john975multiresolution}. These are Discrete Meyer (dmey), Reverse biorthogonal (rbio), Biorthogonal (bior), Daubechies (db), Symlets (sym), Coiflets (coif), and Haar (Haar)~\citep{al2020general}.

\begin{figure}[htpb]
	\centering
	\includegraphics[width=0.7\textwidth]{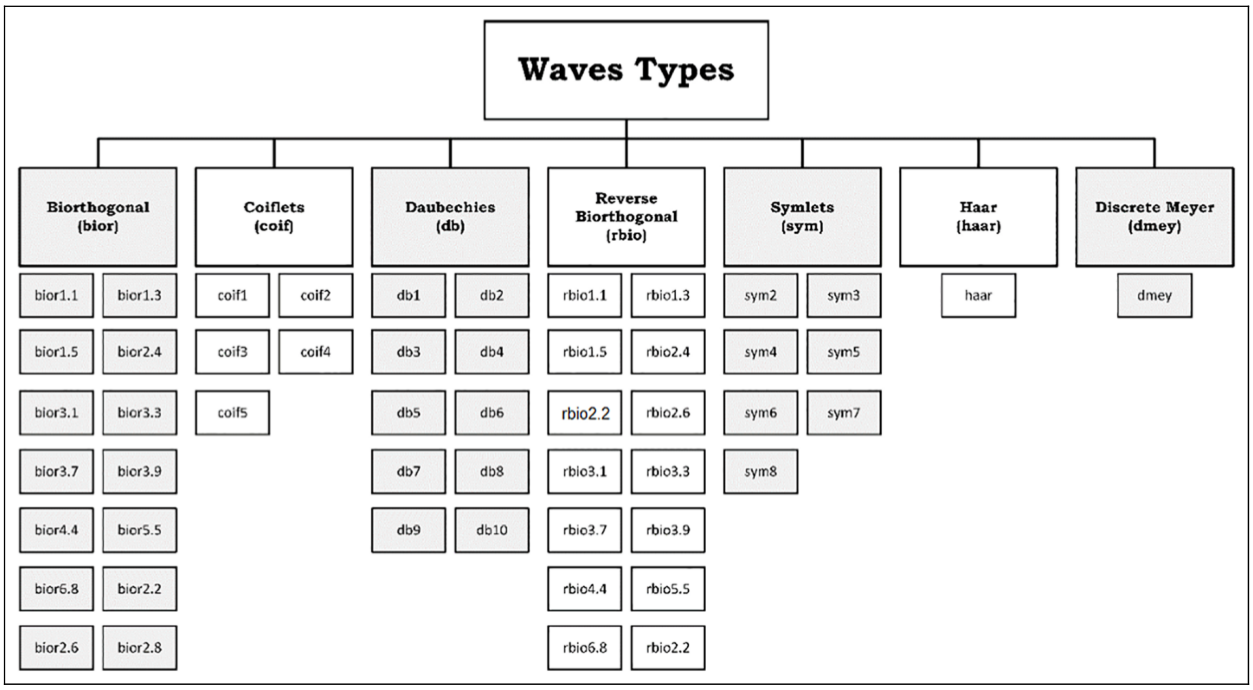}
	\caption{Wavelets Families~\citep{al2020general}} 
	\label{fig:wt}
\end{figure}

Four main factors have a direct impact on discrete wavelet transform (DWT) performance, summarized by: DWT coefficient feature, mother wavelet, frequency band and, decomposition level. As mentioned in~\citep{zhang2017classification}, based on the classification accuracy and computational time, it was found that Coiflet of order 1(Coif1) is the best wavelet family for analysis of EEG signal. According to \citep{john975multiresolution}, this argument is being challenged by many researchers. Therefore they recommend Haar and, second and fourth-order Daubechies (db2, db4) wavelets for signal preprocessing and feature extraction since these methods were provided better accuracy in the recent classifications.The Figure~\ref{fig:wt} presents the different wavelets families and their associated mother wavelet.

\subsubsection{Methodology of DWT}~

The methodology contains three steps which are described below. The coming subsection describes the three critical steps in DWT.
\begin{enumerate}
	\item \textbf{Wavelet Threshold De-Noising :} 
	
	Generally, Physiological signals are nonlinear and contaminated~\citep{wang2017automatic}: it could be due to background noise around the facilities,  disposition of the electrodes, mobility of the patient during the recording, etc. Removing noise is, therefore, an important step. The wavelet threshold method can perform well in denoising non-stationary EEG signals~\citep{john975multiresolution}. The word ``noise" is mentioned as a standard term used in signal processing, but in EEG signal processing noise is in the form of sharp waves that are not significant to identify ~\citep{john975multiresolution}. Thus, getting rid of some frequency bands appearing within the decomposed bands by the use of the wavelet threshold becomes a critical step to achieve to unveil the relevant features from the raw signal. The threshold is expressed as~\citep{gilda2019automatic}:
	\begin{eqnarray}\label{eq32}
	\lambda=\sigma\sqrt{(2\log N)} \ ,
	\end{eqnarray}
	where $\lambda$ is the wavelet threshold, $\sigma$ is the standard deviation of the noise and $N$ is the length of the sample signals, respectively.
	
	The denoised EEG signal will facilitate extraction of distinguishable features than original signal, especially for epileptic event detection~\citep{john975multiresolution}.

	\item \textbf{Wavelet Decomposition}
	
	The process of Wavelet decomposition is described below:
	\begin{itemize}
		\item The first EEG signal goes into a band-pass filter. Band-pass filter is a combination of both High-band-Pass Filter (HPF) and Low-band-Pass Filter (LPF) to have the requested result. This process is categorized as the first level, which incorporates two corresponding coefficients; one among them is Approximation (A) and another is Detailed (D).
		\item The output of the Low-band-Pass Filter (LPF) goes into another band-pass filter and so on up to level 4 as shown in Figure~\ref{fig:decom}.
	\end{itemize}
	Note that, the process goes on under multiple levels as a subsequence of the coefficient from the first level within the approximation. At each process, the frequency resolution is doubled using the filters while decomposing and reducing the time complexity to half.
	\begin{figure}[htbp!]
		\centering 
		\includegraphics[width=0.6\textwidth]{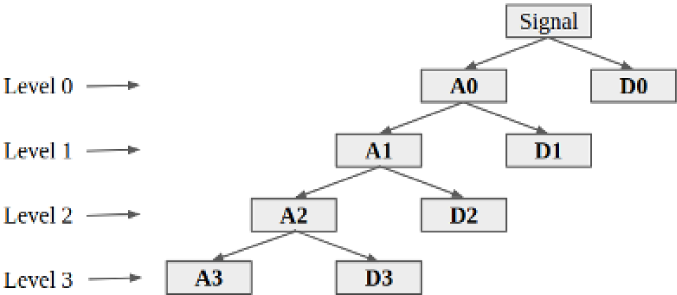}
		\caption{Four level EEG signal decomposition\citep{gilda2019automatic}}
		\label{fig:decom}
	\end{figure}
	
	\item \textbf{Features Extraction :}
	
	Multi-Resolution Analysis (MRA) is used to extract feature vectors from signal data. Commonly, when DWT is used as a feature extraction method, the extracted features for classification include mean average value, standard deviation, energy, and spectral entropy. Below are the formula to compute all these quantities~\citep{ahammad2014detection}.
	\begin{itemize}
		\item The variance can be defined as a deviation of the signal from its mean. It is given as,
		\begin{eqnarray}
		\sigma^2=\dfrac{\sum_{i=1}^{N}(x_i-\mu)^2}{N}
		\end{eqnarray}
		
		\item The mean signal energy of a seizure data generally tends to be higher than that of normal data due to higher amplitudes. It is given as,
		\begin{eqnarray}
		E=\dfrac{\sum_{i=1}^{N}x_i^2}{N}
		\end{eqnarray}
		
		\item Power spectral density is calculated in two steps. First, by finding the fast Fourier transform $X(w_i)$ of the time series and then taking the squared modulus of the FFT coefficients.
		\begin{eqnarray}
		P(w_i)=\dfrac{|X(w_i)|^2}{N}
		\end{eqnarray}
		From this, the maximum and minimum values are used.
		\item Entropy is the measure of randomness and the information content of a signal. To calculate the entropy of a given EEG signal, The Shannon entropy formula is used:
		\begin{eqnarray}
		ENT=-\sum_{i=1}^{N}x_i^2\log(x_i^2)
		\end{eqnarray}
		\item The interquartile range of the EEG signal gives the statistical dispersion of a signal, which is a measure of how squeezed or stretched a distribution is. The signal is divided into two parts, one containing values
		lower than the median and another containing values higher than the median to calculate the inter quartile range. Therefore, it is the difference between the median of the left half and that of the right half.
		
		\item Kurtosis is used to measure  how much weight the tail of the probability distribution has. Hence, the presence of spikes is indicated by increased kurtosis. Kurtosis can be calculated by:
		\begin{eqnarray}
		K=\dfrac{\sum_{i=1}^{N}(x_i-\mu)^4}{N\sigma^4}
		\end{eqnarray}
		Where $x_i$ denotes the time series of an EEG data set and N the number of samples in the signal.
	\end{itemize}
	
\end{enumerate}
The second feature extraction method that we are going to use in our implementation is Mel Frequency Cepstral Coefficient(MFCC) and it is described below.

\subsection{Mel Frequency Cepstral Coefficient}~

\begin{figure}[htbp!]
	\centering 
	\includegraphics[width=0.6\textwidth]{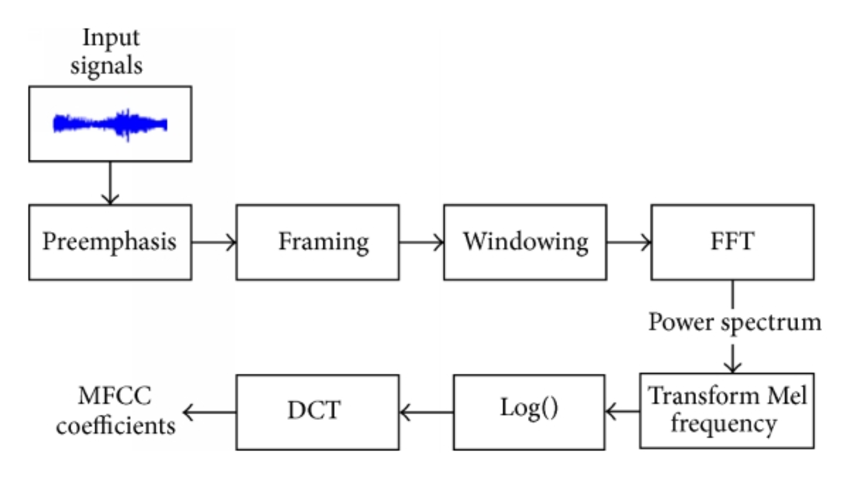}
	\caption{Extraction flowchart of MFCC coefficients~\citep{gong2015emotion}}
	\label{fig:flow_mfcc1}
\end{figure}

MFCC was originally designed to study real speech registered by human ears but it can process quasi- stationary signals in the forms of sound signals and EEG signals~\citep{nguyen2012proposed}. This method is widely used in EEG signal processing with high accuracy~\citep{othman2009mfcc}. The methodology for its implementation is described below (Figure~\ref{fig:flow_mfcc1}).

\subsubsection{Methodology}~

According to~\citep{ren2018cepstrum}, the steps of the MFCC are described as follows:
\begin{enumerate}
	\item \textbf{Pre-emphasis and Framing :}
	Passing of signal through a filter that emphasizes higher frequencies, and thereafter, divide it into $N$ frames.
	\item \textbf{Hamming window :}
	The hamming window is applied to each frame to obtain windowed frames, and can be calculated as follows:
	\begin{eqnarray}
	y(k)=x(k)\times H(k),
	\end{eqnarray}
	Where  $H(k)$ is given by:
	\begin{eqnarray}
	H(k)=a-b\cos\left(\dfrac{2\pi k}{N-1}, \ \ k=0,1,2,...,N-1 \right)
	\end{eqnarray}
	and $N$ is the number of points in a frame, and a, b denotes the parameters of the hamming window (a = 0.54, b = 0.46). $x(k)$ and $y(k)$ are respectively the input and output signal while $H(k)$ is the Hamming window.
	
	\item \textbf{Changing from the time domain to the frequency domain :}
	The output of this step is called a spectrum. The frequency spectrum $F ( w )$ of the $i^{th}$ frame $x_i( n )$ can be calculated using the Fast Fourier Transform. The short-time power spectrum $|F (w)|^2$ can be calculated and filtered by a Mel-filter bank $B_{Mel}$. The mapping relation from a linear frequency f to Mel-frequency $f_{Mel}$ can be calculated as follow:
	\begin{eqnarray}
	f_{Mel}=\delta\ln\left(1+\dfrac{f}{\nu}\right) \ ,
	\end{eqnarray}
	where f and $f_{Mel}$ denote the linear frequency and the Mel-frequency respectively, and $\delta$, $\nu$ are the parameters ($\delta = 2595$,  $\nu = 700$).
	
	\item \textbf{Mel-frequency wrapping :}
	Calculating the log amplitude at the spectrum into the Mel scale by using a triangle filter bank. The output of the short-time power spectrum $|F ( w )|^2$ can be calculated via this
	Mel-filter bank:
	\begin{eqnarray}
	\theta(M_k)=\ln \left[\sum_{k=1}^{N}|F ( w )|^2H_m(k)\right], \ \ m=1,2,3,...,M
	\end{eqnarray}
	Where $H_m(k)$ is the filter bank.
	\item \textbf{Cepstrum :}
	Transforming mel-spectrum coefficient by using Discrete Cosine Transform (DCT) producing fourteen cepstral coefficients for every frame. The Mel-Frequency
	Cepstrum $c ( k )$ can be calculated by applying the updated Inverse Discrete Cosine
	Transform (IDCT) in the Mel-frequency coordinate spectrum, which can be described by the following formula:
	\begin{eqnarray}
	c_k=\sum_{k=1}^{M}M_k\cos\left(\dfrac{n(l-0.5)\pi}{M}\right)\, \ \ n=1,2,...,p
	\end{eqnarray}
	where $p$ is the dimension of MFCC, $c ( k )$ denotes the $k^{th}$ MFCC, and $p$ is less than the number $M$ of Mel-filters.
\end{enumerate}


\section{Classification Techniques and Performance Evaluation}~
Classification is a supervised learning technique of categorizing a given set of data (structured or unstructured) into classes. The main goal of a classification problem is to identify the category/class to which a new data will fall under. In this chapter, seven classifiers will be used, among which Linear and Quadratic Discriminant Analysis, Naives bayes, Support vector Machine, K-Nearest Neighbor, Random Fores, Gradient Boosting. The coming section will present how they work as well as how their performances are assessed.

\subsection{Linear and Quadratic Discriminant Analysis (LDA / QDA)}
Let us assume that, the data set is defined by:
\begin{equation}
\{(x_i,y_i)\}_{i=1}^{n} \ ,
\end{equation}
where n is the sample size, $x_i$ and $y_i$ represent respectively the deferent feature vectors and the class labels. For simplicity, we note that $x_i\in \mathbb{R}^{d}$ and $y_i \in \mathbb{R}$. The objectives of the following method is to classify the data into k classes using  Linear Discriminant Analysis (LDA), and Quadratic discriminant Analysis (QDA).

To better understand how these two methods work, let us start our study by considering the one-dimensional data which means $x \in \mathbb{R}$ and just two classes.

Let us denote by $G_1(x)$ and $G_2(x)$ the Cumulative Distribution Functions of these two classes. we can derive the corresponding probability density functions by:
\begin{eqnarray}
g_1(x)=\dfrac{\partial G_1(x)}{\partial x}\\
g_2(x)=\dfrac{\partial G_2(x)}{\partial x}
\end{eqnarray}
Let us assume that the two classes have normal (Gaussian) distribution~\citep{ghojogh2019linear} and that the mean of one of the two classes is greater than the other one $(\mu_1<\mu_2)$. In the following, $\mathcal{C}_1$ and $\mathcal{C}_2$ represent respectively the first and second class.

Let us denote by $x^{*}$, the point where the probability of the two classes are equal. We can write that $\mu_1<x^{*}< \mu_2$ since we know that $ \mu_1< \mu_2$. This means that:
\begin{equation*}
\begin{cases}
if  \  x< x^{*} \ \text{then} \ x  \ \text{belong to} \  \mathcal{C}_1\\
if  \  x> x^{*} \ \text{then} \ x  \ \text{belong to} \  \mathcal{C}_2
\end{cases}
\end{equation*} 

The probability $P_1$ of the error in the estimation of the class where x belongs to can be written as: 
\begin{eqnarray}\label{eq4}
P_1=P(x > x^{*} , x \in  \mathcal{C}_1) + P(x < x^{*} ,x \in  \mathcal{C}_2 )
\end{eqnarray}
Using the conditional probability given by \ref{eqq4}:
\begin{equation}\label{eqq4}
P(A,B)=P(A|B)P(B)
\end{equation}
we can rewrite ~(\ref{eq4}) as :
\begin{equation}\label{eq6}
P_1=P(x > x^{*} | x \in  \mathcal{C}_1)P( x \in  \mathcal{C}_1) + P(x < x^{*} |x \in  \mathcal{C}_2 )P(x \in  \mathcal{C}_2 )
\end{equation}
The Aim of these methods is to minimize $P_{1}$ by finding $x^{*}$.

Using the definition of the Cumulative Distribution Function, we can write:
\begin{eqnarray}\label{eq7}
\begin{cases}
P(x > x^{*} | x \in  \mathcal{C}_1)=1- G_1(x^{*})\\
P(x < x^{*} |x \in  \mathcal{C}_2 )= G_2(x^{*})
\end{cases}
\end{eqnarray}
Using the definition of probability density function, we can write:
\begin{eqnarray}\label{eq8}
\begin{cases}
P( x \in  \mathcal{C}_1)= g_1(x)=\sigma_1\\
P( x \in  \mathcal{C}_2)= g_2(x)=\sigma_2
\end{cases}
\end{eqnarray}
Here we denote the piors $g_1(x)$ and $g_2(x)$ by $\sigma_1$ and $\sigma_1$.
By replacing ~(\ref{eq7}) and ~(\ref{eq8}) in ~(\ref{eq6}), we obtain:
\begin{equation}
P_1=\left[1- G_1(x^{*})\right]\sigma_1+ G_2(x^{*})\sigma_2
\end{equation}
Let us now take the derivative of $P_1$ for the sake of minimization
\begin{eqnarray*}
	\dfrac{\partial P_1}{\partial x^{*}}=-g_1(x^{*})\sigma_1+ g_2(x^{*})\sigma_2
\end{eqnarray*}
If we set this derivative equal to zero, we will obtain the relation:
\begin{eqnarray} \label{eq10}
g_1(x^{*})\sigma_1= g_2(x^{*})\sigma_2
\end{eqnarray}
$g_i(x^{*})$ and $\sigma_1$ are the likelihood (class conditional) and
prior probabilities respectively.
Now let us suppose that the data is a multivariate data with dimensionality d. The probability density function for multivariate Gaussian distribution is given by:
\begin{eqnarray}\label{eq11}
g(x)=\dfrac{1}{\sqrt{(2\pi)^{d}|\Sigma|}}\exp\left(-\dfrac{(x-\mu)^{T}\Sigma^{-1}(x-\mu)}{2}\right) \ ,
\end{eqnarray}
where $x \ \in \ \mathbb{R}^{d}$, $\mu \ \in \ \mathbb{R}^{d}$ is the mean, $\Sigma \ \in \ \mathbb{R}^{d\times d}$ is the covariance matrix, and $|.|$ is the determinant of the matrix.
Replacing ~(\ref{eq11}) in ~(\ref{eq10}), we obtain:
\begin{eqnarray}\label{eq12}
\dfrac{1}{\sqrt{(2\pi)^{d}|\Sigma|_1}}\exp\left(-\dfrac{(x_1-\mu_1)^{T}\Sigma_1^{-1}(x_1-\mu_1)}{2}\right)\sigma_1 = \dfrac{1}{\sqrt{(2\pi)^{d}|\Sigma|_2}}\nonumber\\
\times \exp\left(-\dfrac{(x_1-\mu_2)^{T}\Sigma_2^{-1}(x_1-\mu_2)}{2}\right)\sigma_2
\end{eqnarray}
This last equation is used for LDA and QDA methods
\subsubsection{Linear Discriminant Analysis for Binary Classification(LDA)}~

In this case, we assume that the two classes have equal covariance matrices $(\Sigma_1=\Sigma_2=\Sigma)$ \cite{ghojogh2019linear}. Therefore, the equation ~(\ref{eq12}) becomes
\begin{eqnarray*}
\exp\left(-\dfrac{(x_1-\mu_1)^{T}\Sigma^{-1}(x_1-\mu_1)}{2}\right)\sigma_1=\exp\left(-\dfrac{(x_1-\mu_2)^{T}\Sigma^{-1}(x_1-\mu_2)}{2}\right)\sigma_2
\end{eqnarray*}
By taking the logarithm of both sizes we obtain:
\begin{eqnarray}\label{eq14}
-\dfrac{(x_1-\mu_1)^{T}\Sigma^{-1}(x_1-\mu_1)}{2}+\ln(\sigma_1)=-\dfrac{(x_1-\mu_2)^{T}\Sigma^{-1}(x_1-\mu_2)}{2}+\ln(\sigma_2)
\end{eqnarray} 
The numerator of the first term of the left-hand side can be developed and rewritten as:
\begin{eqnarray}\label{eq15}
(x_1-\mu_1)^{T}\Sigma^{-1}(x_1-\mu_1)=x_1^{T}\Sigma^{-1}x_1+\mu_1^{T}\Sigma^{-1}\mu_1-2\mu_1^{T}\Sigma^{-1}x_1
\end{eqnarray}
because \ $\mu_1^{T}\Sigma^{-1}x_1= x_1^{T}\Sigma^{-T} \mu_1$ \ and \  $\Sigma^{-T}=\Sigma^{-1}$ since $\Sigma^{-1}$ is symmetric.
We can do the same with the numerator of right hand side and obtain:
\begin{eqnarray}\label{eq16}
(x_1-\mu_2)^{T}\Sigma^{-1}(x_1-\mu_2)=x_1^{T}\Sigma^{-1}x_1+\mu_2^{T}\Sigma^{-1}\mu_2-2\mu_2^{T}\Sigma^{-1}x_1
\end{eqnarray}
By replacing ~(\ref{eq15}) and ~(\ref{eq16}) in ~(\ref{eq14}) and arranging we obtain:
\begin{eqnarray}
2\left(\Sigma^{-1}(\mu_2-\mu_1)\right)^{T}x_1 + (\mu_1-\mu_2)^{T}\Sigma^{-1}(\mu_1-\mu_2) + 2\ln(\dfrac{\sigma_2}{\sigma_1})
\end{eqnarray}
This last equation is the equation of a line. Thus, if we consider Gaussian distributions for the two classes where the covariance matrices are assumed to be equal, the boundary of classification is a line. Because of this linearity which discriminates the two classes, the method is called Linear Discriminant Analysis (LDA).

If we define $\Gamma(x)$ : $\mathbb{R}^{d} \longmapsto \mathbb{R}$ such that: 
\begin{eqnarray}
\Gamma(x_1)&=&2\left(\Sigma^{-1}(\mu_2-\mu_1)\right)^{T}x_1 + (\mu_1-\mu_2)^{T}\Sigma^{-1}(\mu_1-\mu_2) + 2\ln(\dfrac{\sigma_2}{\sigma_1})
\end{eqnarray}
then the class of an instance x is estimated as:
\begin{eqnarray}\label{eq19'}
\mathcal{C}(x)=
\begin{cases}
1 \ if \ \Gamma(x)< 0\\
2 \ if \ \Gamma(x)> 0
\end{cases}
\end{eqnarray}

\subsubsection{Quadratic Discriminant Analysis for Binary Classification(QDA)}~

In this case, we do not assume that the two classes have equal covariance matrices, so we have $(\Sigma_1\neq\Sigma_2 )$ \citep{ghojogh2019linear}. Therefore, by taking the natural logarithm of both sides of equation ~(\ref{eq12}) we obtain:
\begin{eqnarray}\label{eq18}
-\dfrac{1}{2}\ln (|\Sigma_1|)-\dfrac{(x_1-\mu_1)^{T}\Sigma_1^{-1}(x_1-\mu_1)}{2}+\ln(\sigma_1)=-\dfrac{1}{2}\ln (|\Sigma_2|) \nonumber\\
-\dfrac{(x_1-\mu_2)^{T}\Sigma_2^{-1}(x_1-\mu_2)}{2}+\ln(\sigma_2)
\end{eqnarray}
According to ~(\ref{eq15}), we can rewrite ~(\ref{eq18}) as:
\begin{eqnarray}\label{eq19}
-\dfrac{1}{2}\ln (|\Sigma_1|)-\dfrac{1}{2}x_1^{T}\Sigma_1^{-1}x_1-\dfrac{1}{2}\mu_1^{T}\Sigma_1^{-1}\mu_1+ \mu_1^{T}\Sigma_1^{-1}x_1 + \ln(\sigma_1) \nonumber\\
=-\dfrac{1}{2}\ln(|\Sigma_2|)-\dfrac{1}{2}x_1^{T}\Sigma_2^{-1}x_1-\dfrac{1}{2}\mu_2^{T}\Sigma_2^{-1}\mu_2+ \mu_2^{T}\Sigma_2^{-1}x_1 + \ln(\sigma_2)
\end{eqnarray}
Let us multiply ~(\ref{eq19}) by 2. After some rearrangement, we obtain:
\begin{eqnarray*}
x_1^{T}\left(\Sigma_1-\Sigma_2\right)^{-1}x_1 + 2 \left(\Sigma_2^{-1}\mu_2- \Sigma_1^{-1}\mu_1\right)^{T}x_1 + \left(\mu_1^{T}\Sigma_1^{-1}\mu_1 - \mu_2^{T}\Sigma_2^{-1}\mu_2\right) \nonumber\\ +\ln\left(\dfrac{|\Sigma_1|}{|\Sigma_2|}\right) +2\ln\left(\dfrac{\sigma_2}{\sigma_1}\right)=0
\end{eqnarray*}
which is in the quadratic form. Hence, if we consider Gaussian distributions for the two classes, the decision boundary of classification is quadratic, that is why this method is called Quadratic Discriminant Analysis(QDA).

Let us define $\Gamma(x)$ : $\mathbb{R}^{d} \longmapsto \mathbb{R}$ such that: 
\begin{eqnarray}
\Gamma(x_1)&=&x_1^{T}\left(\Sigma_1-\Sigma_2\right)^{-1}x_1 + 2 \left(\Sigma_2^{-1}\mu_2- \Sigma_1^{-1}\mu_1\right)^{T}x_1 + \left(\mu_1^{T}\Sigma_1^{-1}\mu_1 - \mu_2^{T}\Sigma_2^{-1}\mu_2\right) \nonumber\\ &+&\ln\left(\dfrac{|\Sigma_1|}{|\Sigma_2|}\right) +2\ln\left(\dfrac{\sigma_2}{\sigma_1}\right)
\end{eqnarray}
For estimation of the class of an instance x, we use equation ~(\ref{eq19'}).

\subsubsection{LDA and QDA for Multi-Class Classification}~

In this general case, we consider multiple classes (they can be more than two) indexed by k $\in$ $\{1, . . . , |C|\}$.
According to Equations ~(\ref{eq10}) and ~(\ref{eq11}) we have:
\begin{eqnarray}\label{eq24}
g_k(x)\sigma_k= \dfrac{1}{\sqrt{(2\pi)^{d}|\Sigma_k|}}\exp\left(-\dfrac{(x-\mu_k)^{T}\Sigma_k^{-1}(x-\mu_k)}{2}\right)\sigma_k
\end{eqnarray}
If we take the logarithm of ~(\ref{eq24}), we have:
\begin{eqnarray*}
\ln (g_k(x)\sigma_k)=-\dfrac{d}{2}\ln(2\pi)-\dfrac{1}{2}\ln(|\Sigma_k|)- \dfrac{1}{2}(x-\mu_k)^{T}\Sigma_k^{-1}(x-\mu_k)+\ln(\Sigma_k|)
\end{eqnarray*}
Let us drop the first term because it is the same for all the classes. So we have:
\begin{eqnarray*}
\Gamma_k(x)=-\dfrac{1}{2}\ln(|\Sigma_k|)- \dfrac{1}{2}(x-\mu_k)^{T}\Sigma_k^{-1}(x-\mu_k)+\ln(|\Sigma_k|)
\end{eqnarray*}
$\Gamma_k(x)$ is the scaled posterior of the k-th class.

In QDA, the class of instance x is estimated as:
\begin{eqnarray}\label{eq27}
\hat{C}(x)= \underset{k}{\arg\max} \ \Gamma_k(x)
\end{eqnarray}

In LDA we assume that $\Sigma_1=\Sigma_2=...=\Sigma_{|C|}=\Sigma$. Therefore $\Gamma_k(x)$ becomes:
\begin{eqnarray*}
\Gamma_k(x)=-\dfrac{1}{2}\ln(|\Sigma|)-\dfrac{1}{2}x^{T}\Sigma^{-1}x-\dfrac{1}{2}\mu_k^{T}\Sigma^{-1}\mu_k + \mu_k^{T}\Sigma^{-1}x + \ln(\sigma_k)
\end{eqnarray*}
We drop the first and second of the right-hand side term because it is the same for all the classes. Hence, we have:
\begin{eqnarray*}
\Gamma_k(x)= \mu_k^{T}\Sigma^{-1}x-\dfrac{1}{2}\mu_k^{T}\Sigma^{-1}\mu_k + \ln(\sigma_k)
\end{eqnarray*}
So the class of the instance x is determined by ~(\ref{eq27})

\subsection{Naive Bayes}
Naive Bayes classifier is a classifier based on the Bayes Theorem with the naive assumption that, given their belonging to a specific class, features are independent of each other. Let's see how we can derive the model.
\begin{defn}
	Bayes Theorem~\citep{stuart1994kendall}
	
Given a feature vector $X=(x_1,x_2,...,x_n)$ and a class variable $C_k$, the Bayes Theorem states that:
\begin{eqnarray}\label{eq431}
P(C_k|X)=\dfrac{P(X|C_k)P(C_k)}{P(X)},\ \mbox{ for}n\ k=1,2,...,K \ ,
\end{eqnarray}
\end{defn}
$P(C_k|X)$ is called the posterior probability, $P(X|C_k)$ the likelihood, $P(C_k)$ the prior probability of class and $P(X)$ the prior probability of predictor.

We are interested in calculating the posterior probability from the likelihood and prior probabilities. Using the chain rule, the likelihood $P(X|Ck)$ can be decomposed as:
\begin{eqnarray}\label{eq432}
P(X|C_k)=P(x_1|x_2,...x_n,C_k)P(x_2|x_3,...x_n,C_k)...P(x_{n-1}|x_n,C_k)P(x_n|C_k)
\end{eqnarray}
The above sets of probabilities can be hard and expensive to compute. However we can use the Naive independence assumption which is given by: 
\begin{eqnarray}\label{eq433}
P(x_i\mid x_{i+1},...,x_n\mid C_k) = P(x_i\mid C_k)
\end{eqnarray}
Using (~\ref{eq433}) in (~\ref{eq432}), we have:
\begin{eqnarray}
P(X\mid C_k) = P(x_1,...,x_n\mid C_k) = \prod_{i=1}^n{P(x_i\mid C_k)}
\end{eqnarray}
Therefore, the posterior probability (~\ref{eq431}) can then be written as:
\begin{eqnarray}\label{eq435}
P(C_k|X) = \frac{P(C_k)\prod_{i=1}^n{P(x_i\mid C_k)}}{P(X)}
\end{eqnarray}
Knowing that the prior probability of predictor $P(X)$ is constant given the input, we can write (\ref{eq435}) as:
\begin{eqnarray}
P(C_k|X) \propto P(C_k)\prod_{i=1}^n{P(x_i\mid C_k)}
\end{eqnarray}
where $\propto$ means positively proportional to.

The Naive Bayes classification problem is : for different class values of Ck, find the maximum of$$P(C_k)\prod_{i=1}^n{P(x_i\mid C_k)}$$. Mathematically, we can formulate this problem as:
\begin{eqnarray}
\hat{C} = \arg \max_{C_k} P(C_k)\prod_{i=1}^n{P(x_i\mid C_k)}
\end{eqnarray}
The prior probability of class $P(C_k)$ could be estimated as the relative frequency of class $C_k$ in the training data.

\subsubsection{Advantages of Naive Bayes}
\begin{itemize}
	\item When the assumption of independent predictors holds true, a Naive Bayes classifier performs better as compared to other models.
	\item Naive Bayes requires a small amount of training data to estimate the test data. So the training period takes less time.
	\item It can be used for both binary and multi-class classification problems.
\end{itemize}
\subsubsection{Disadvantages of Naive Bayes}
\begin{itemize}
	\item The main limitation of Naive Bayes is the assumption of independent predictor features. Naive Bayes implicitly assumes that all the attributes are mutually independent. In real life, it is almost impossible that we get a set of predictors that are completely independent or one another.
	\item If a categorical variable has a category in the test dataset, which was not observed in the training dataset, then the model will assign a 0 (zero) probability and will be unable to make a prediction.
\end{itemize}
\subsection{Support Vector Machine}
Support Vector Machine (SVM) is a supervised machine learning algorithm that can be used for both classification or regression problems.  In the SVM algorithm, we plot each data item as a point in n-dimensional space (where n is number of features ), with the value of each feature being the value of a particular coordinate. Then, we perform classification by finding the hyperplane that differentiates the classes very well. It can be use for binary classification and multi-class classification.Below we present the mathematics behind binary classification.

Let us assume that, the dataset is define by:
\begin{equation}
\{(x_i,y_i)\}_{i=1}^{n} \ ,
\end{equation}
where n is the sample size, $x_i$ and $y_i$ represent respectively the different feature vectors and the class labels. We note that $x_i\in \mathbb{R}^{d}$ and $y_i \in \{-1,1\}$.

According to~\citep{nguyen2012proposed}, Support Vector Machine (SVM) algorithm will find the optimal hyperplane given by~(\ref{eqq43}):
\begin{eqnarray}\label{eqq43}
f (x) = w^T\Phi(x) + b
\end{eqnarray}
to separate the training data by solving the optimization problem ~(\ref{eq43})
\begin{eqnarray}\label{eq43}
min \dfrac{1}{2}\parallel w\parallel^2+\mathcal{C}\sum_{i=1}^{n}\psi_i
\end{eqnarray}
subject to the constraint ~(\ref{eq44}).
\begin{eqnarray}\label{eq44}
y_i (w^T\Phi(x_i) + b) \geq 1 -\psi_i \ \mbox{and} \ \psi_i \geq 0 , i = 1, ..., n
\end{eqnarray}
The optimization problem ~(\ref{eq43}) will guarantee to maximize the hyperplane margin while minimizing the cost of error, where $\psi_i , i = 1, ..., n$ are non-negative slack variables introduced to relax the constraints of separable data problems to the constraint ~(\ref{eq44}) of non-separable data problems. For an error to occur, the corresponding must exceed unity, so $\sum_{i}^{}\psi_i$ is an upper bound on the number of training errors. Hence an extra cost $\mathcal{C}_i\psi_i$ for errors is added to the objective function where $\mathcal{C}$ is a parameter chosen by the user.

For Nonlinear classification (case where, the data cannot be separated by an hyperplane),  the kernel function $K(x_i , x_j ) = \Phi(x_i)^T \Phi(x_j)$ is introduced and the optimal hyperplane becomes:
\begin{eqnarray}
	f(x)= \sum_{i=1}^n\alpha_iy_iK(s_i,x )+b \ ,
\end{eqnarray}
where $s_i$ is the $i^{th}$ support vector. The function $\Phi : x^{(i)} \longrightarrow \Phi(x)^{(i)}$ is a map from the data space to the feature space such that the data are linearly separable in the feature space.

Note that there are several kernel functions such as Polynomial kernel, Gaussian kernel and Sigmoidal Kernel.

\subsection{K-Nearest Neighbor}
K-Nearest Neighbor (KNN) is the simplest classification algorithm. The approach is to plot all data points on space, and with any new sample, observe its k nearest points on space and make a decision based on majority voting. Thus, KNN algorithm involves no training and it takes the least calculation time when implemented with an optimal value of k. The steps of KNN algorithm are as follows~\cite{sutton2012introduction}:

\begin{itemize}
	\item  For a given instance, find its distance from all other data points. Use an appropriate distance
	metric based on the problem instance.
	\item  Sort the computed distances in increasing order. Depending on the value of k, observe the nearest
	k points.
	\item Identify the majority class amongst the k points, and declare it as the predicted class.
\end{itemize}
Choosing an optimal value of k is a challenge in this approach. Most often, the process is repeated for several different trials of k. The evaluation scores are then observed using a graph to find the optimal value of k.

\subsection{Random Forest}
Random forests can be define as a combination of tree predictors such that each tree depends on the values of a random vector sampled independently and with the same distribution for all trees in the forest. The steps of building a random forest classifier are as follows~\citep{cutler2012random}:
\begin{itemize}
	\item Select a subset of features from the dataset.
	\item From the selected subset of features, using the best split method, pick a node.
	\item Continue the best split method to form child nodes from the subset of features.
	\item Repeat the steps until all nodes are used as split.
	\item Iteratively create n numbers of trees using steps 1-4 to form a forest.
\end{itemize}
\begin{figure}[htbp!]
	\centering 
	\includegraphics[width=0.69\textwidth]{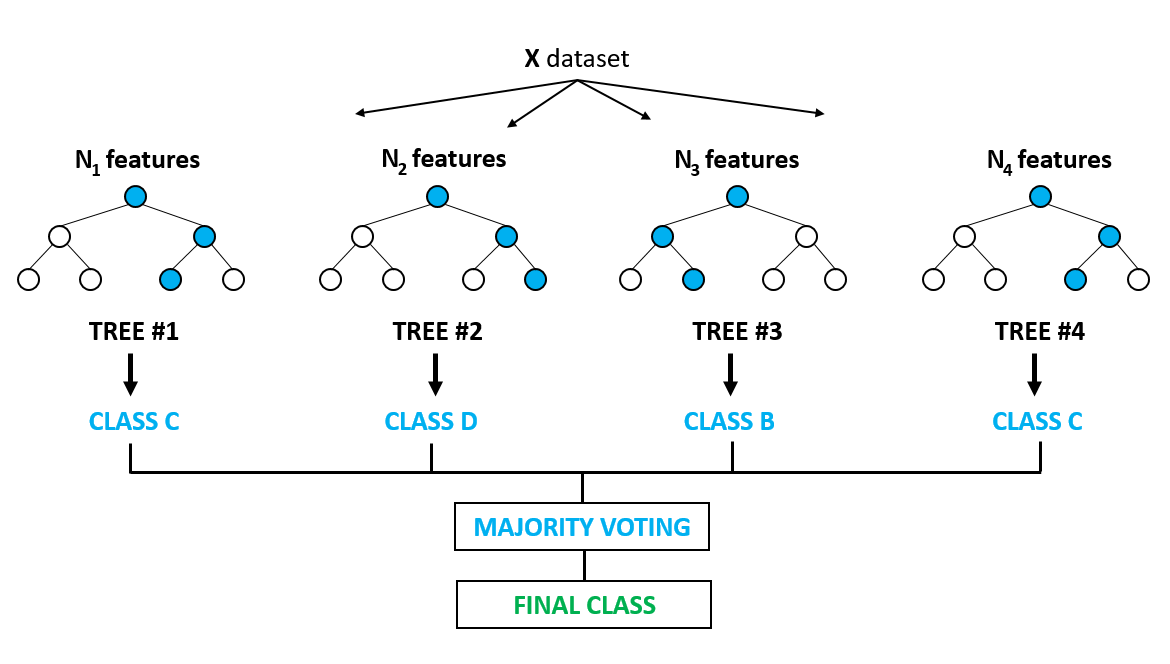}
	\caption{Steps of Random Forest~\citep{cutler2012random}}
	\label{fig:RF}
\end{figure}
\subsection{Gradient Boosting}
Gradient Boosting is one of the technique for performing supervised machine learning tasks, like classification and regression. Like Random Forests, it is an ensemble learner. This means that, it will create a final model based on a collection of individual models. Let us see how it works. The magic of this model is described in the name: ``Gradient" plus ``Boosting".

Boosting built models from individual in an iterative way. In boosting, the individual models are not built on completely random subsets of data and features but sequentially by putting more weight on instances with wrong predictions and high errors. The general idea behind this is that instances, which are hard to predict correctly will be focused on during learning, so that the model learns from past mistakes. When we train each ensemble on a subset of the training set, it is called Stochastic Gradient Boosting, which can help improve generalizability of the model~\citep{natekin2013gradient}.

Similar to how Neural Networks utilize gradient descent to optimize ("learn") weights, the gradient is used to minimize a loss function. The weak learner is built and its predictions are compared to the correct outcome that we expect in each round of training. The error rate of the model is estimated by the distance between prediction, and truth which can be used to calculate the gradient. The gradient is basically the partial derivative of the loss function, thus it describes the steepness of the error function~\citep{natekin2013gradient}.
\subsection{Performance Evaluation}
To train and evaluate the performance of different models, we will use K-fold cross-validation with 5 replications followed by the confusion matrix which is describe below.
\subsubsection{K-fold cross-validation}

Cross-validation, also know as rotation estimation, is the statistical practice of partitioning a sample of data into subsets such that the analysis is initially performed on a single subset, while the other subset(s) are retained for subsequent use in confirming and validating the initial analysis. The initial subset of data is called the training set; the other subset(s) are called validation or testing sets. In K-fold cross-validation, the original sample is partitioned into K sub-samples. Between K sub-samples, a single subsample is retained as the validation data for testing the model, and the remaining $K-1$ sub samples are used as training data. The cross-validation process is then repeated K times (the folds), with each of the K sub-samples used exactly once as the validation data. The K results from the folds then can be averaged to produce a single estimation.

\begin{itemize}
	\item The \textbf{advantage} of this method over repeated random sub-sampling is that all observations are used for both training and validation, and each observation is used for validation exactly once. The variance of the resulting estimate is reduced as k is increased.
	\item The \textbf{disadvantage} of this method is that the training algorithm has to be run again from scratch k times, which means it takes k times as much computation to make an evaluation.
\end{itemize}

\subsubsection{Metrics evaluation}
For performance recognition evaluation, confusion matrix metrics are often used. The criteria for performance evaluation usually employed include three parts: \textbf{sensitivity} (the proportion of the total number of positive cases that are correctly classified), \textbf{specificity} (the proportion of the total number of negative cases that are correctly classified), and classification \textbf{accuracy} (the proportion of the total number of EEG signals that are correctly classified).

\section{Experimental Results and Analysis}
\begin{figure}[h]
	\centering
	\includegraphics[width=0.8\textwidth]{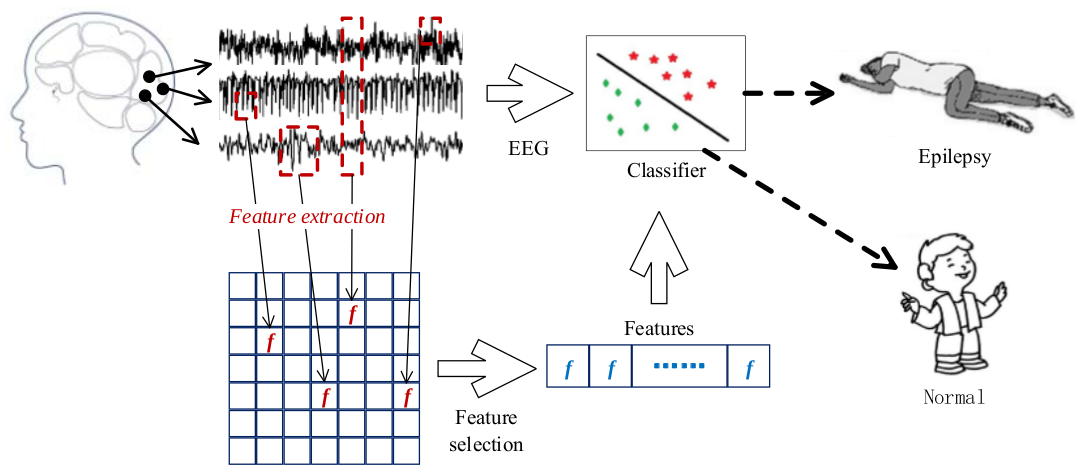}
	\caption{Process of EEG signal classification~\citep{wen2017effective}}
	\label{fig:class_pro} 
\end{figure}

~
In this section, we implement the different methods presented above on the EEG dataset. As features extraction methods, we implement Discrete Wavelet Transform (DWT) and Mel Frequency Ceptral Coefficient (MFCC) and as a classification method, we implement QDA, LDA, RF, NB, GB, KNN, SVM. We also make a comparison not only between the different feature extraction methods but also the interaction between feature extraction and the classifier based on the final scores of the classifiers. The overal process of classification is describeb by Figure~\ref{fig:class_pro}.
\section{Methodology}
\begin{figure}[h]
	\centering
	\includegraphics[width=1\textwidth]{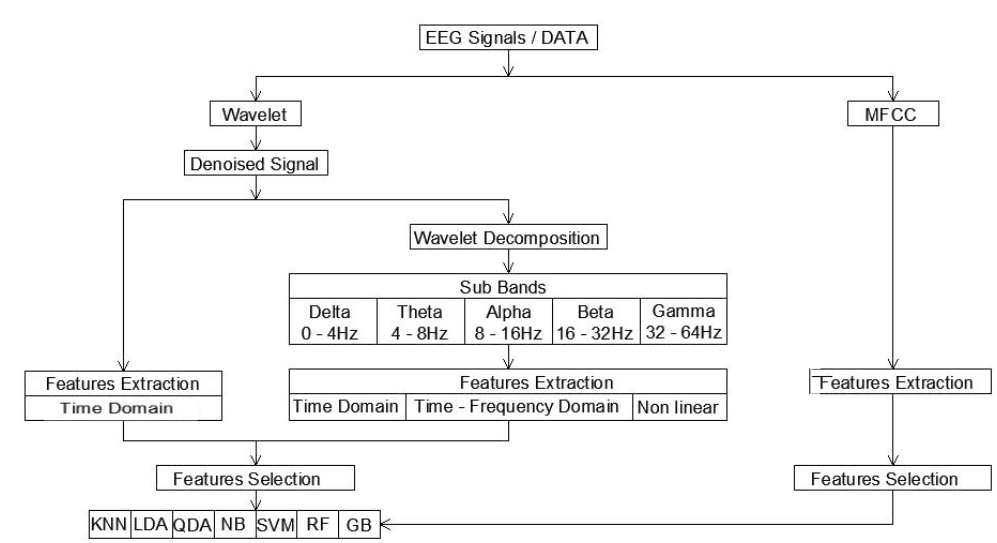}
	\caption{The flowchart of the proposed classification framework}
	\label{fig:flow_method} 
\end{figure}

~
The flowchart of the proposed classification framework is shown by Figure~\ref{fig:flow_method}

\subsection{Description of Database}~

The database used in this work is recorded at the University Hospital Bonn, Germany. It is composed of five different sections of EEG signals, and these sections are represented by symbols S, F, N, O, and Z as shown in Table~\ref{tab:data}. Each of these sections consist of 100 signals, where recording time was about 23.6 s. In order to record the data in the most accurate way, they used an amplified system with 128 signal channels, in which the output resulted in  173.61 Hz of the sampling rate. The EEG samples in the O and Z datasets are derived from healthy volunteers with external surface electrodes for open and closed eye conditions. The F and N datasets are acquired during seizure-free intervals and the dataset S contains only the seizure activity. The five data sets S, F, N, O, and Z are classified into two distinct groups in our study. The epileptic seizure class (S) is composed of the subset S, and the non-seizure class (FNOZ) is composed of the subsets F, N, O, and Z, respectively.
\begin{table}[htbp]
	\centering
	\caption{The definitions and descriptions for the electroencephalographic (EEG) signals from the University of Bonn, Germany}
	\begin{tabular}{|p{0.15\linewidth}|p{0.16\linewidth}|p{0.16\linewidth}|p{0.12\linewidth}|p{0.12\linewidth}|p{0.13\linewidth}|}
		\hline 
		\textbf{Information}& \textbf{Dataset O}&\textbf{Dataset Z} &\textbf{Dataset F}&\textbf{Dataset N}&\textbf{Dataset S}  \\ 
		\hline 
		State &Awake and eyes open(Healthy) & Awake and eyes closed(Healthy) &Seizure-free &Seizure-free &Seizure activity \\ 
		\hline 
		Electrode type &Surface &Surface & Intercranial & Intercranial & Intercranial  \\ 
		\hline 
		$N_o$ of channels &100 &100  &100 &100  &100 \\ 
		\hline 
		Recording time &23.6  &23.6    &23.6   & 23.6   &23.6   \\ 
		\hline 
	\end{tabular}  
	\label{tab:data}
\end{table}
	
\subsection{Data Preparation}~

\begin{figure}[h]
	\centering
	\includegraphics[width=0.9\textwidth]{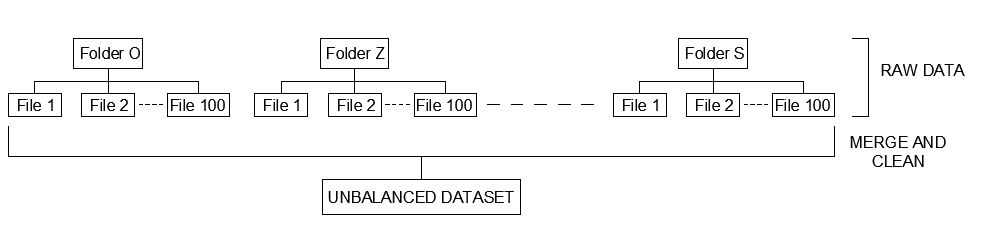}
	\caption{Flowchart for the building of the imbalanced dataset}
	\label{fig:flow_unbalanced} 
\end{figure}

\begin{figure}[h]
	\centering 
	\includegraphics[width=0.9\textwidth]{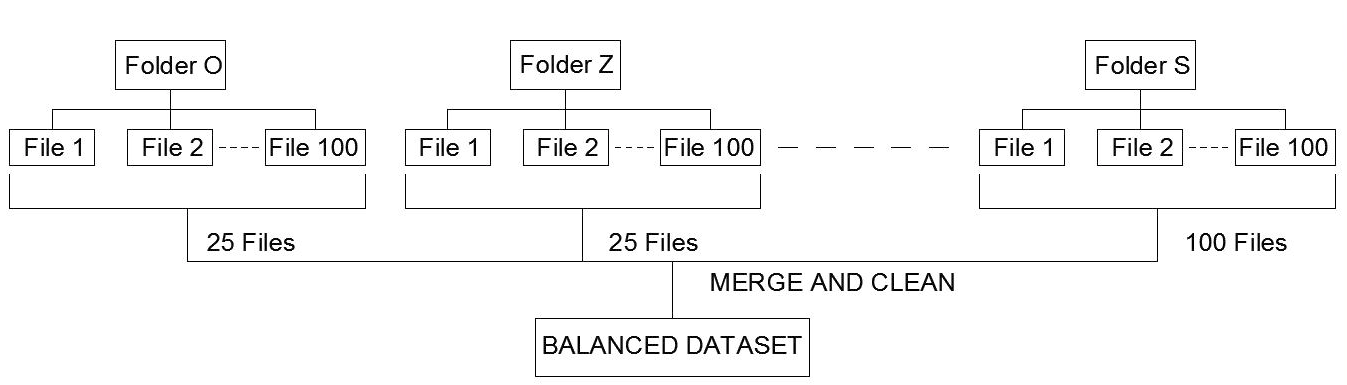}
	\caption{Flowchart for the building of the balanced dataset}
	\label{fig:flow_balanced}
\end{figure}

In this study, we built and used two datasets from the raw database. The first dataset is an imbalanced dataset with 0.2 as the prevalence of the positive class (Figure~\ref{fig:flow_unbalanced}) and the second is a balanced dataset with 0.5 as the prevalence of the positive class (Figure~\ref{fig:flow_balanced}).

\subsection{Feature Engineering}~

\begin{figure}[h]
	\centering
	\includegraphics[width=0.8\textwidth]{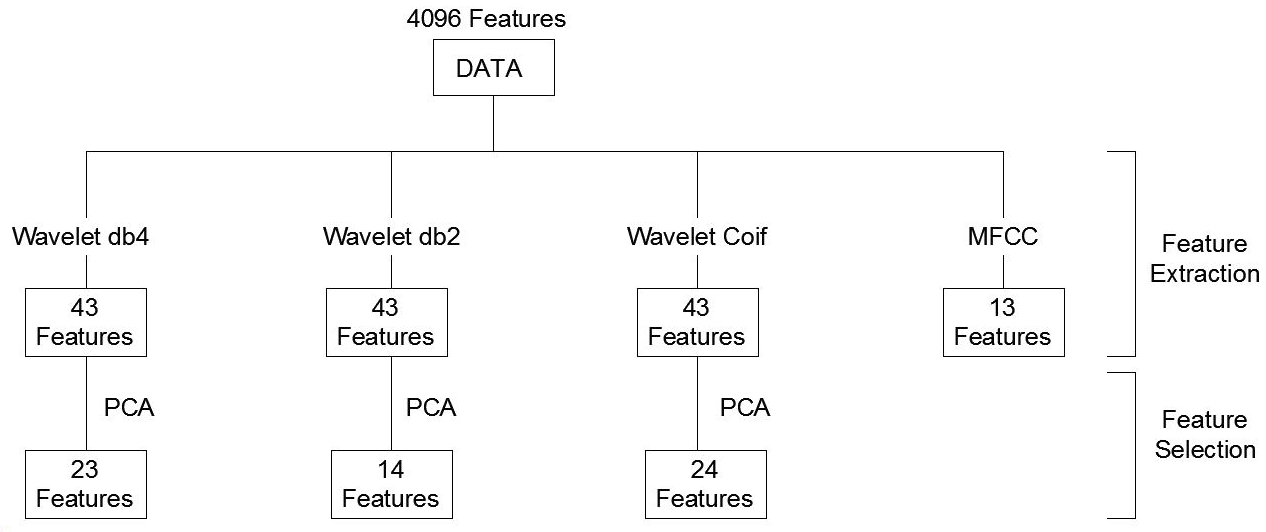}
	\caption{The flowchart of Feature Engineering}
	\label{fig:flow_feat} 
\end{figure}

Feature engineering is the process of converting raws data into features that better represent the raw observations for predictive models. Thus, it can improve the accuracy of the model on unseen data. The flowchart of our feature engineering is described in Figure~\ref{fig:flow_feat}. 

\subsubsection{Features extraction}~

\begin{figure}[h]
	\centering 
	\includegraphics[width=0.9\textwidth]{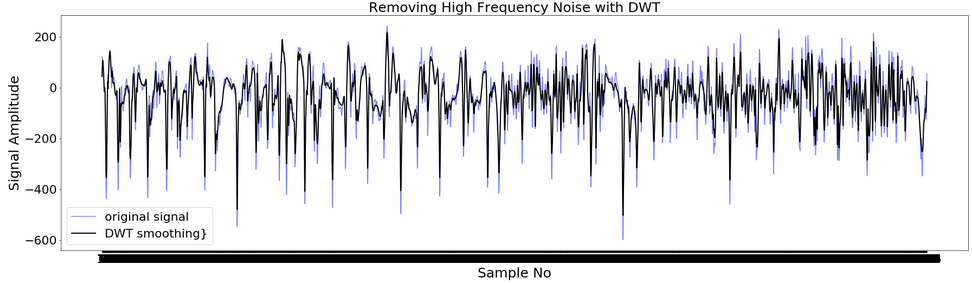}
	\caption{Signal before and after de-noising }
	\label{fig:denois}
\end{figure}

\begin{figure}[h]
	\centering 
	\includegraphics[width=0.9\textwidth]{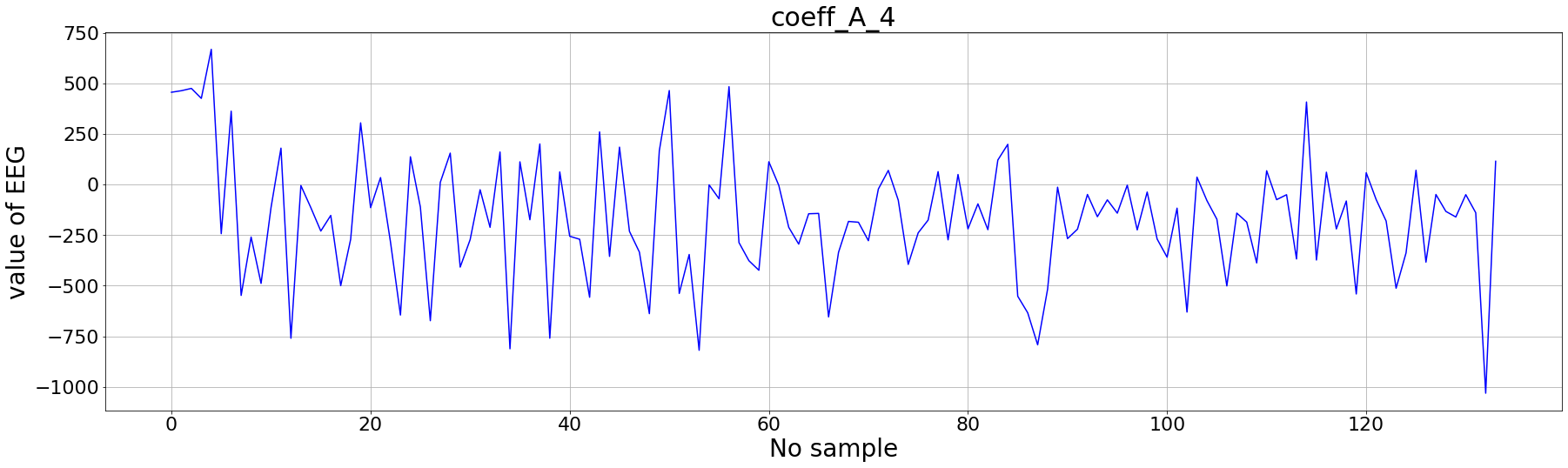}
	\caption{Sub-band signals using four-level wavelet decomposition (db4) from an original EEG signal: the approximation in 0-4 Hz.}
	\label{fig:wav_a4}
\end{figure}

\begin{figure}[h]
	\centering 
	\includegraphics[width=0.9\textwidth]{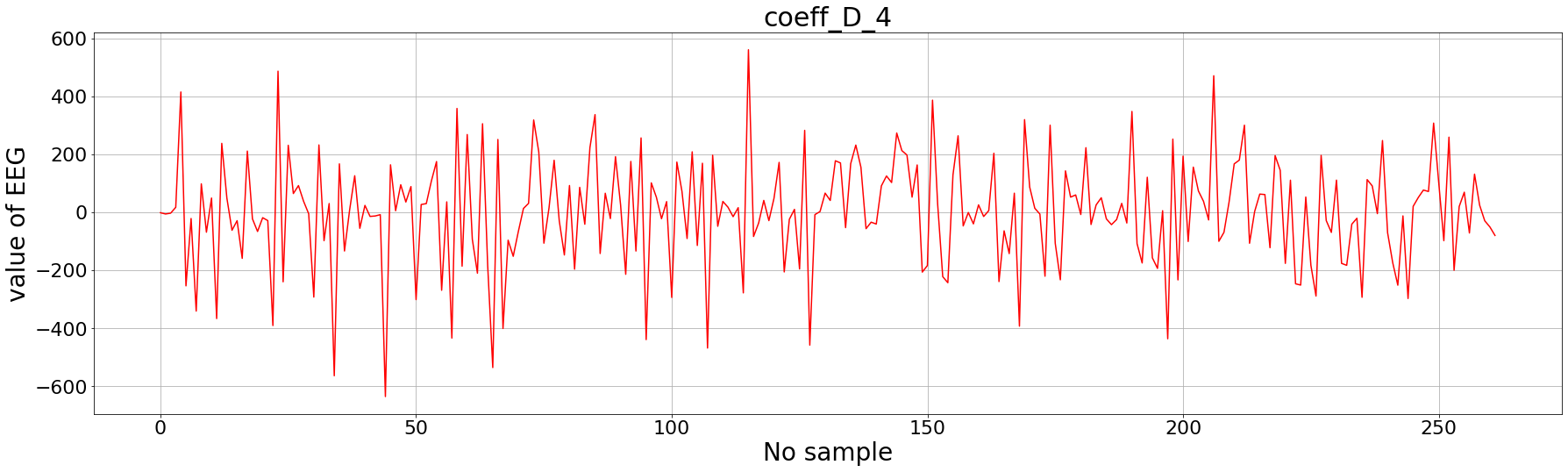}
	\caption{Sub-band signals using four-level wavelet decomposition (db4) from an original EEG signal: the detail in 4-8 Hz.}
	\label{fig:wav_d4}
\end{figure}

\begin{figure}[h]
	\centering 
	\includegraphics[width=0.9\textwidth]{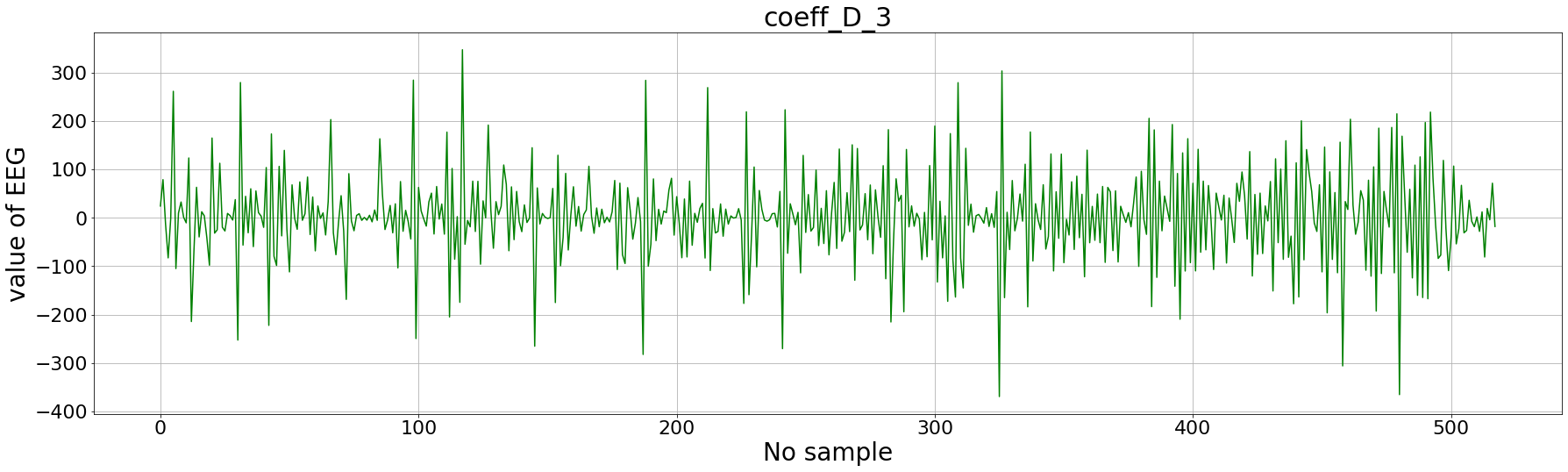}
	\caption{Sub-band signals using four-level wavelet decomposition (db4) from an original EEG signal: the detail in 8-16 Hz.}
	\label{fig:wav_d3}
\end{figure}

\begin{figure}[h]
	\centering 
	\includegraphics[width=0.9\textwidth]{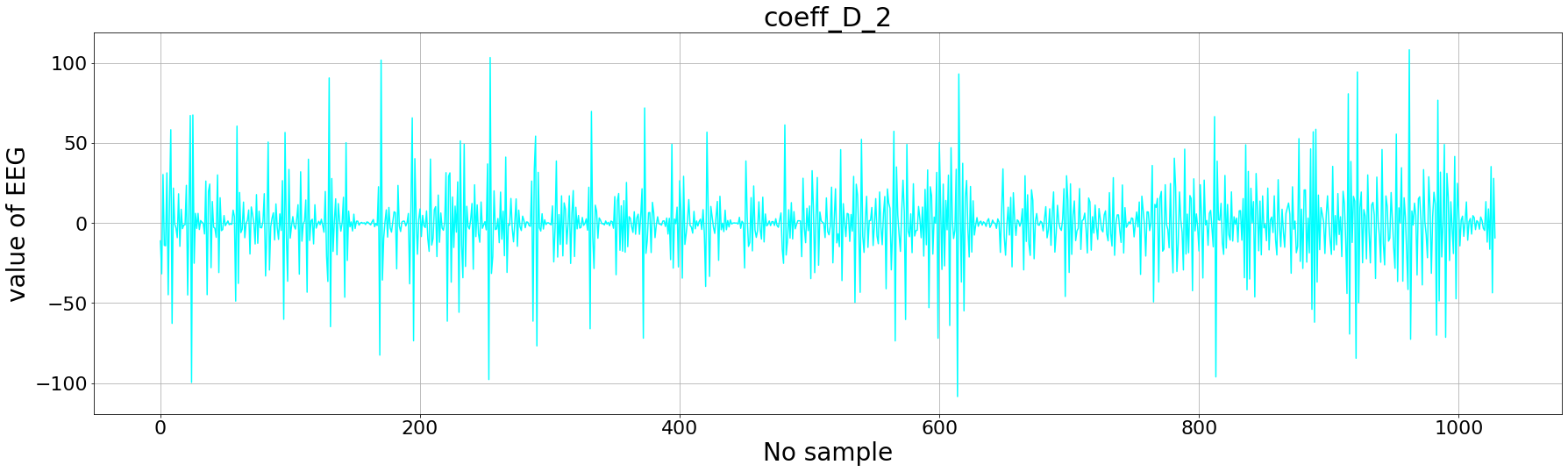}
	\caption{Sub-band signals using four-level wavelet decomposition (db4) from an original EEG signal: the detail in 16-32 Hz. }
	\label{fig:wav_d2}
\end{figure}

\begin{figure}[h]
	\centering 
	\includegraphics[width=0.9\textwidth]{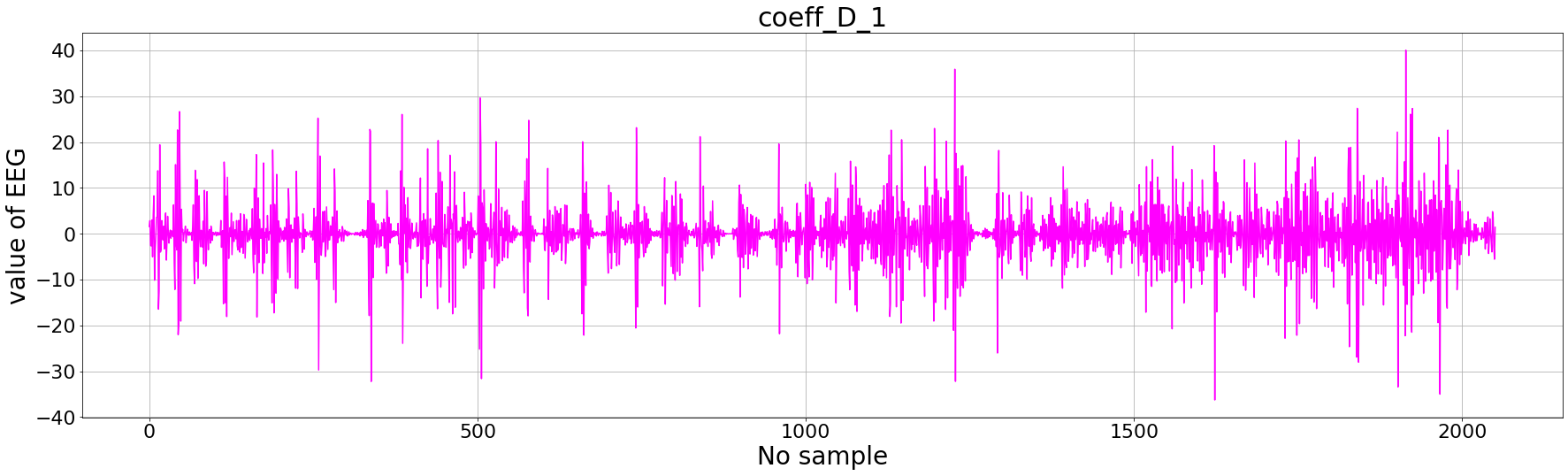}
	\caption{Sub-band signals using four-level wavelet decomposition (db4) from an original EEG signal. the detail in 32-64 Hz. }
	\label{fig:wav_d1}
\end{figure}

For features extraction, we have implemented 4 feature extraction methods namely :
\begin{itemize}
	\item Discrete wavelet transform DWT-db4,
	
	\item Discrete wavelet transform DWT-db2,
	
	\item Discrete wavelet transform DWT-coif1, 
	
	\item Mel Frequency Ceptral Coeficient MFCC. 
\end{itemize}
We follow the step of wavelet transform for the first three and the step of MFCC describeb in chapter three for the last one.
\begin{itemize}
	\item \textbf{Wavelet Threshold De-Noising :} We can see in Figure~\ref{fig:denois} one signal after and before de-noising 
\end{itemize}
\begin{itemize}
	\item \textbf{Wavelet Decomposition}: The DWT is used to split a signal into different frequency sub-bands, as many as needed or as many as possible. We can see in figures~(\ref{fig:wav_a4},~\ref{fig:wav_d4},~\ref{fig:wav_d3},~\ref{fig:wav_d2},~\ref{fig:wav_d1}) the five decomposed bands from one signal that we are going to use.
\end{itemize}

\begin{itemize}
	\item \textbf{Feature Extraction}: After decomposition, we extract some features in the time-frequency Domain (mean average value,standard deviation, Relative band power, spectral entropy), in the frequency Domain (relative power spectral density estimated by the coefficients of the FFT), and in time domain (
	mean, median, standard deviation and the total variation). Additionally, the maximum, minimum and the total variation measures of the DWT transform coefficients are also estimated in order to describe the non-stationary signals.
\end{itemize}


\subsubsection{Features selection}~

Some of the original features extracted are correlated and redundant. To select an optimal feature subset from the original feature set, we use one feature dimensionality reduction method named (PCA). The PCA algorithm is implemented to obtain a relatively low dimensional, but significantly discriminative feature set which will improve the classification performance

\subsection{Classification}~

Seven classifiers are used after 4 feature extraction methods to detect the epileptic seizure from the non-seizure.
The differents classifier are:
\begin{itemize}
	\item Linear discriminant analysis (LDA),
	
	\item Quadratic discriminant analysis (QDA),
	
	\item K-nearest neigbor (KNN),
	
	\item Naive Bayes (NB),
	
	\item Random forest (RF),
	
	\item Gradient boosting (GB),
	
	\item Suport vector machine (SVM).
\end{itemize}
The results of the classification performance measured by the Sensitivity (SEN), the Specificity (SPE), and Accuracy (ACC) using 10-fold cross-validation are shown in the different tables. Besides, boxplot are used to compare the different models and features extraction methods.

\subsubsection{Classification Results}~

\begin{table}[htbp]
	\centering
	\caption{The classification performance for 10-fold CV without features extraction}
	\begin{tabular}{p{0.10\linewidth}p{0.10\linewidth}p{0.10\linewidth}p{0.10\linewidth}p{0.10\linewidth}p{0.10\linewidth}p{0.10\linewidth}}
		\hline 
		& \multicolumn{3}{c}{\textbf{10-Fold cv}}&\multicolumn{3}{c}{\textbf{10-Fold cv}} \\ 
		\hline 
		\textbf{Classifiers}& \textbf{ACC(\%)}&\textbf{SPE(\%)} &\textbf{SEN(\%)}&\textbf{ACC(\%)}&\textbf{SPE(\%)} &\textbf{SEN(\%)} \\ 
		\hline 
		LDA &69.60 &94.50 &40.00 &86.80 &94.00 &44.00\\ 
		QDA &92.90 &99.75 &86.00 &\textbf{96.92}  &\textbf{100 } &\textbf{85.00}\\ 
		KNN &75.50 &100 &42.00 &88.48 &100 &52.00 \\ 
		NB &94.7 &96.50  &88.00 &95.16  &99.00 &90.00\\ 
		RF &94.30  &99.00  &81.00 & 95.44 &95.00 &96.00 \\ 
		GB &\textbf{95.1} &\textbf{95.36} &\textbf{82.00} &96.00  &98.00 &94.00 \\ 
		SVM & 92.30&95.50  &85.00 &96.60 &99.00 &86.00 \\  
		\hline 
		& \multicolumn{3}{c}{\textbf{Imbalanced data}}&\multicolumn{3}{c}{\textbf{Balanced data}}\\
		\hline
	\end{tabular} 
	\label{tab:not}
\end{table}

 
\begin{table}[htbp]
	\centering
	\caption{The classification performance of 10-fold CV with wavelet the ``db4"method}
	\begin{tabular}{p{0.10\linewidth}p{0.10\linewidth}p{0.10\linewidth}p{0.10\linewidth}p{0.10\linewidth}p{0.10\linewidth}p{0.10\linewidth}}
		\hline 
		& \multicolumn{3}{c}{\textbf{10-Fold cv}}&\multicolumn{3}{c}{\textbf{10-Fold cv}} \\ 
		\hline 
		\textbf{Classifiers}& \textbf{ACC(\%)}&\textbf{SPE(\%)} &\textbf{SEN(\%)}&\textbf{ACC(\%)}&\textbf{SPE(\%)} &\textbf{SEN(\%)} \\
		\hline 
		LDA &98.50 & 99.25&83.00&95.24 &99.00 &99.00\\ 
		QDA & 96.40& 99.25&87.00&97.04 &100.00 &94.00\\ 
		KNN & 97.90&99.50 &86.00 &97.00&100.00 &94.00 \\ 
		NB &91.80 &93.75 &87.00 &92.24 &99.00 &95.00\\ 
		RF &\textbf{98.90} &\textbf{99.1}& \textbf{98.00}&97.76  &98.00 &100.00 \\ 
		GB & 98.60& 99.00&97.00 &98.28  &98.00 &99.00 \\ 
		SVM &\textbf{98.91}&\textbf{99.1} &\textbf{98.00}&\textbf{98.99} &\textbf{99.00} &\textbf{99.00} \\ 
		\hline 
		& \multicolumn{3}{c}{\textbf{Imbalanced data}}&\multicolumn{3}{c}{\textbf{Balanced data}}\\
		\hline
	\end{tabular} 
	\label{tab:db4}
\end{table}



\begin{table}[htbp]
	\centering
	\caption{The classification performance of 10-fold CV with wavelet the ``db2"method}
	\begin{tabular}{p{0.10\linewidth}p{0.10\linewidth}p{0.10\linewidth}p{0.10\linewidth}p{0.10\linewidth}p{0.10\linewidth}p{0.10\linewidth}}
		\hline 
		& \multicolumn{3}{c|}{\textbf{10-Fold cv}}&\multicolumn{3}{c}{\textbf{10-Fold cv}} \\ 
		\hline 
		\textbf{Classifiers}& \textbf{ACC(\%)}&\textbf{SPE(\%)} &\textbf{SEN(\%)}&\textbf{ACC(\%)}&\textbf{SPE(\%)} &\textbf{SEN(\%)} \\
		\hline 
		LDA & 96.10&99.25 &89.00&95.60&100.00&98.00\\ 
		QDA &94.9&97.25 & 91.00&97.84& 99.00&91.00\\ 
		KNN &94.80 &97.50 &84.00 &96.92 &100.00&89.00 \\ 
		NB &92.70 & 94.00 &84.00 &91.88 &93.00&92.00\\ 
		RF &96.20&99.50 &85.00 & \textbf{98.56}&\textbf{98.10}&\textbf{97.00} \\ 
		GB &96.10&98.75 &93.00 & 98.48 &98.00&97.00 \\ 
		SVM &\textbf{98.99}&\textbf{ 99.25}&\textbf{95.00} &98.40 &99.00&99.00 \\ 
		\hline 
		& \multicolumn{3}{c}{\textbf{Imbalanced data}}&\multicolumn{3}{c}{\textbf{Balanced data}}\\
		\hline 
	\end{tabular} 
	\label{tab:db2}
\end{table}



\begin{table}[htbp]
	\centering
	\caption{The classification performance of 10-fold CV with wavelet the ``coif1"method}
	\begin{tabular}{p{0.10\linewidth}p{0.10\linewidth}p{0.10\linewidth}p{0.10\linewidth}p{0.10\linewidth}p{0.10\linewidth}p{0.10\linewidth}}
		\hline 
		& \multicolumn{3}{c}{\textbf{10-Fold cv}}&\multicolumn{3}{c}{\textbf{10-Fold cv}} \\ 
		\hline 
		\textbf{Classifiers}& \textbf{ACC(\%)}&\textbf{SPE(\%)} &\textbf{SEN(\%)}&\textbf{ACC(\%)}&\textbf{SPE(\%)} &\textbf{SEN(\%)} \\
		\hline 
		LDA & 94.20&99.50&88.00 &95.28 &100.00 &96.00\\ 
		QDA & 93.30&98.50&90.00 &97.52 &99.00 &91.00\\ 
		KNN & 95.30&99.75&82.00 & 96.04&100.00&92.00\\ 
		NB & 92.80 &95.00 &84.0 &92.40&91.00&95.00\\ 
		RF & 95.90 &98.50 &92.00 &97.04&95.00 &98.00 \\ 
		GB &96.20 &98.25&94.00 &97.20 &95.00&98.25\\ 
		SVM &\textbf{97.80}&\textbf{99.25}&\textbf{93.00} &\textbf{97.92} &\textbf{98.00}&\textbf{98.00}\\  
		\hline 
		& \multicolumn{3}{c}{\textbf{Imbalanced data}}&\multicolumn{3}{c}{\textbf{Balanced data}}\\
		\hline
	\end{tabular} 
	\label{tab:coif1}
\end{table}



\begin{table}[htbp]
	\centering
	\caption{The classification performance of 10-fold CV with the MFCC method}
	\begin{tabular}{p{0.10\linewidth}p{0.10\linewidth}p{0.10\linewidth}p{0.10\linewidth}p{0.10\linewidth}p{0.10\linewidth}p{0.10\linewidth}}
		\hline 
		& \multicolumn{3}{c}{\textbf{10-Fold cv}}&\multicolumn{3}{c}{\textbf{10-Fold cv}} \\ 
		\hline 
		\textbf{Classifiers}& \textbf{ACC(\%)}&\textbf{SPE(\%)} &\textbf{SEN(\%)}&\textbf{ACC(\%)}&\textbf{SPE(\%)} &\textbf{SEN(\%)} \\
		\hline 
		LDA &93.70 &99.00 &82.00 &95.24 &97.00 &90.00\\ 
		QDA &94.00 &99.05 &83.00 &94.00 &95.00  &93.00\\ 
		KNN &90.70 &98.25 &78.00 &94.20 &95.50 &93.00 \\ 
		NB &92.30 &96.75  &85.00 &94.72 &95.00 &94.00\\ 
		RF &94.40 &98.50  &77.00 &95.80 &96.50 &94.00 \\ 
		GB &\textbf{95.60 }&\textbf{96.50} &\textbf{85.00} &94.48 &95.00 &92.00 \\ 
		SVM &94.10 &99.00  &83.00 &\textbf{96.80} &\textbf{97.00} &\textbf{94.00} \\ 
		\hline 
		& \multicolumn{3}{c}{\textbf{Imbalanced data}}&\multicolumn{3}{c}{\textbf{Balanced data}}\\
		\hline
	\end{tabular} 
	\label{tab:mfcc}
\end{table}

\subsection{Models Comparison}~

\begin{figure}[htbp!]
	\centering
	\includegraphics[width=0.9\textwidth]{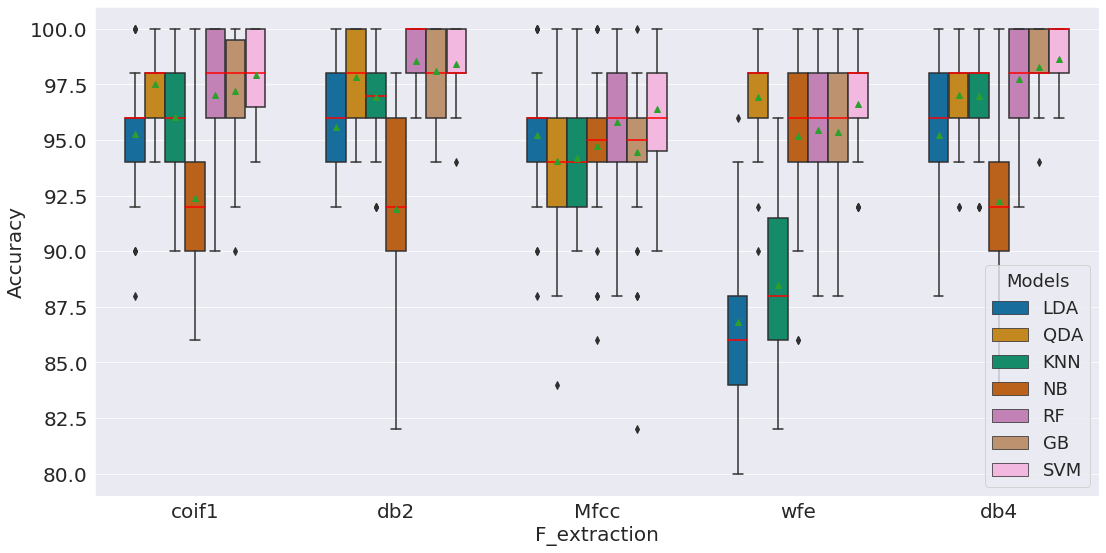}
	\caption{Box-plot for model comparison in terms of accuracy in the balanced data}
	\label{fig:comp_balanced} 
\end{figure}

\subsubsection{Case of the balanced dataset}~

According to Figure~\ref{fig:comp_balanced}, we can draw the conclusion below for the balanced dataset:

\begin{itemize}
	\item Without features extraction :
	\begin{itemize}
		\item The best model is Quadratic Discriminant Analysis (QDA) which has 96.92\% of accuracy, 100\% of specificity, and 85\% of sensitivity .
		\item LDA and KNN have the smallest accuracy and can be avoided when we performed a classification without features extraction.
	\end{itemize}
	\item Among the feature extraction methods :
	\begin{itemize}
		\item  Wavelet db4 is the best method when it is used with SVM which has 98.99\% of Accuracy 99\% of sensitivity and 99\% of specificity.
		\item Wavelet db2 associated with RF challenge db4 associated with SVM with 98.56\% of Accuracy, 97\% of sensitivity, and 98.10\% of specificity.
		\item Wavelet coif1 brings its best results when it is Associated with SVM and the results are 97.92\% of Accuracy, 98\% of sensitivity, and 98\% of specificity.
		\item MFCC performs less than all the DWT used here but can bring its best results when associated to SVM.
	\end{itemize}
\end{itemize}


\begin{figure}[htbp!]
	\centering
	\includegraphics[width=0.9\textwidth]{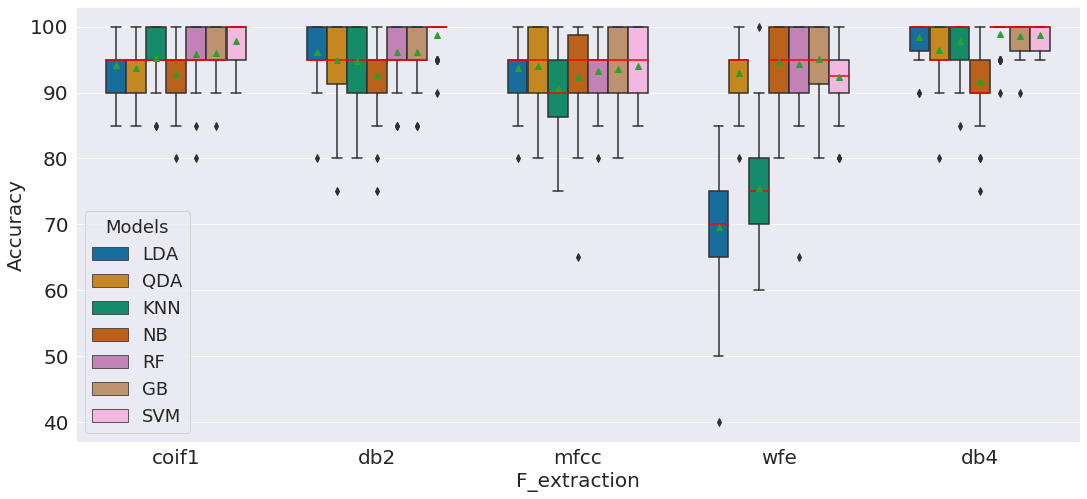}
	\caption{ Box-plot for model comparison in term of accuracy in the imbalanced dataset}
	\label{fig:comp_unbalanced} 
\end{figure}

\subsubsection{Case of the imbalanced dataset}~

According to Figure~\ref{fig:comp_unbalanced}, we can draw the conclusion below for the imbalanced dataset:
\begin{itemize}
	\item Without features extraction :
	\begin{itemize}
		\item The best model is GB which has 95.1\% of accuracy, 82.00\% of sensitivity and 95.36\% of specificity.
		\item LDA and KNN have the smallest accuracy and can be avoided when we performed a classification without features extraction.
	\end{itemize}
	\item Among the feature extraction methods :
	\begin{itemize}
		\item Wavelet db2 is the best method when it is associated with RF which brings 98.99\% of accuracy, 99.25\% of specificity, and 95\% of sensitivity.  
		
		\item  Wavelet db4 can challenge wavelet db2 when it is used with SVM or RF which here bring 98.90\% of accuracy, 99.10\% of specificity, and 98\% of sensitivity.
		
		\item Wavelet coif1 brings its best results when it is associated with SVM and the results are 97.80\% of accuracy, 99.25\% of specificity, and 93\% of sensitivity.
		\item MFCC performs less than all the DWT used here but can bring its best results when it is associated with SVM.
	\end{itemize}
\end{itemize}
Looking at the results presented above, we observe many differences in the predictive performances. In the next part, we check the statistical significance of the difference in predictive performances and the effect Size.

\subsection{Statistical Significance of the Difference in the Predictive Performances}~

There are two factors to evaluate; \textbf{feature extraction method} and \textbf{model}. They have five and seven levels respectively. Therefore, the two-way Analysis Of Variance (ANOVA) is suitable for our analysis. Using the two-way ANOVA, we can simultaneously evaluate how the type of feature extraction and model affect the accuracy of classification. Hence, we can test three effects below on classification accuracy:

\begin{itemize}
	\item Effect of features extraction.
	\item Effect of models.
	\item Effect of features extraction and models interactions.
\end{itemize}

\subsubsection{Case of the imbalanced dataset}~

From Table~\ref{Tab:Anova}, the P-value obtained from ANOVA analysis for feature extraction, models, and interaction are statistically significant $(P\leq0.05)$. We conclude that, the type of feature extraction, the type of model, and interaction of both feature extraction and model significantly affects the accuracy of classification.

\begin{table}[h]
	\centering
	\caption{ANOVA analysis for the imbalanced dataset}
	\begin{tabular}{lrrrrr}
		\hline
		& Df & Sum Sq & Mean Sq & F value & Pr($>$F) \\ 
		\hline
		Models           & 6 & 8947.34 & 1491.22 & 60.10 & 0.0000 \\ 
		feat\_extr        & 4 & 18879.49 & 4719.87 & 190.23 & 0.0000 \\ 
		Models:feat\_extr & 24 & 28916.51 & 1204.85 & 48.56 & 0.0000 \\ 
		Residuals        & 1715 & 42552.50 & 24.81 &  &  \\ 
		\hline
	\end{tabular}
	\label{Tab:Anova}
\end{table}

Each factor has an independent significant effect on classification accuracy. While it is good to know if there is a statistically significant effect of some models or feature extraction techniques on the accuracy, it is as important to know the size of the effect they have on the outcome. To check this, we can calculate the effect size which is estimated by the measures of omega-squared presented in Table~\ref{Tab:omega-squared_I}. 

\begin{table}[h]
	\centering
	\caption{omega-squared}
	\begin{tabular}{rlr}
		\hline
		& names & $\Omega^{2}$ \\ 
		\hline
		1 & Models           & 0.09 \\ 
		2 & feat\_extr        & 0.19 \\ 
		3 & Models:feat\_extr & 0.29 \\ 
		\hline
	\end{tabular}
	\label{Tab:omega-squared_I}
\end{table}

They are an estimate of how much variance in the response variables is accounted for by the explanatory variables.
The following interpretation of  omega-squared are suggested by~\citep{field2013discovering}.
\begin{itemize}
	\item Omega-squared = 0 - 0.01: very small
	\item Omega-squared = 0.01 - 0.06: small
	\item Omega-squared = 0.06 - 0.14: medium
	\item Omega-squared $>$ 0.14: large
\end{itemize} 
According to this, Model has a medium effect on the mean accuracy while feature extraction and the interaction between model and feature extraction have a large effect. 

\subsubsection{Case of the balanced dataset}~

From Table~\ref{Tab:Anova2}, the P-value obtained from ANOVA analysis for feature extraction, models, and interaction are statistically significant $(P\leq0.05)$. We conclude that, the type of feature extraction, the type of model and, interaction of both feature extraction and model significantly affect the accuracy of classification.

\begin{table}[h]
	\centering
	\caption{ANOVA analysis for the balanced dataset}
	\begin{tabular}{lrrrrr}
		\hline
		& Df & Sum Sq & Mean Sq & F value & Pr($>$F) \\ 
		\hline
		Models           & 6 & 4607.28 & 767.88 & 111.95 & 0.0000 \\ 
		feat\_extr        & 4 & 2564.63 & 641.16 & 93.48 & 0.0000 \\ 
		Models:feat\_extr & 24 & 4953.55 & 206.40 & 30.09 & 0.0000 \\ 
		Residuals        & 1715 & 11762.96 & 6.86 &  &  \\ 
		\hline
	\end{tabular}
	\label{Tab:Anova2}
\end{table}

As previously seen, each factor has an independent significant effect on the classification accuracy. For the effect size, model, and the interaction between model and feature extraction have a large effect on the mean accuracy while feature extraction has a medium effect (Table~\ref{Tab:omega-squared_b}).

\begin{table}[h]
	\centering
	\caption{omega-squared}
	\begin{tabular}{rlr}
		\hline
		& names & $\Omega^{2}$ \\ 
		\hline
		1 & Models           & 0.19 \\ 
		2 & feat\_extr        & 0.11 \\ 
		3 & Models:feat\_extr & 0.20 \\ 
		\hline
	\end{tabular}
	\label{Tab:omega-squared_b}
\end{table}

Now, given that the feature extraction methods alongside the suitable models are statistically significant upon a certain effect size, it is important to mention that, ANOVA does not tell much about which of the two models outperform the other. To know the pairs of significant different feature extraction methods, and type of model, we can perform a multiple pairwise comparison analysis.

\subsubsection{Statistical Significance of the Difference in Predictive Performances of MFCC}~

\begin{table}[h]
	\centering
	\caption{HSD-test for pairwise comparison of features extraction (Imbalanced dataset)}
	\begin{tabular}{rllrrrr}
		\hline
		& term & comparison & estimate & conf.low & conf.high & adj.p.value \\ 
		\hline
		1 & feat\_extr & mfcc-wfe & 5.31 & 4.29 & 6.34 & 0.00 \\ 
		2 & feat\_extr & coif1-wfe & 7.33 & 6.30 & 8.36 & 0.00 \\ 
		3 & feat\_extr & db2-wfe & 7.89 & 6.86 & 8.91 & 0.00 \\ 
		4 & feat\_extr & db4-wfe & 9.49 & 8.46 & 10.51 & 0.00 \\ 
		5 & feat\_extr & coif1-mfcc & 2.01 & 0.99 & 3.04 & 0.00 \\ 
		6 & feat\_extr & db2-mfcc & 2.57 & 1.54 & 3.60 & 0.00 \\ 
		7 & feat\_extr & db4-mfcc & 4.17 & 3.14 & 5.20 & 0.00 \\ 
		8 & feat\_extr & db2-coif1 & 0.56 & -0.47 & 1.59 & 0.58 \\ 
		9 & feat\_extr & db4-coif1 & 2.16 & 1.13 & 3.19 & 0.00 \\ 
		10 & feat\_extr & db4-db2 & 1.60 & 0.57 & 2.63 & 0.00 \\ 
		\hline
	\end{tabular}
	\label{fig:hsd_I}
\end{table}

\begin{table}[h]
	\centering
	\caption{HSD-test for pairwise comparison of features extraction (Balanced dataset)}
	\begin{tabular}{rllrrrr}
		\hline
		& term & comparison & estimate & conf.low & conf.high & adj.p.value \\ 
		\hline
		1 & feat\_extr & Mfcc-wfe & 1.45 & 0.91 & 1.99 & 0.00 \\ 
		2 & feat\_extr & coif1-wfe & 2.66 & 2.12 & 3.20 & 0.00 \\ 
		3 & feat\_extr & db4-wfe & 3.06 & 2.52 & 3.60 & 0.00 \\ 
		4 & feat\_extr & db2-wfe & 3.22 & 2.68 & 3.76 & 0.00 \\ 
		5 & feat\_extr & coif1-Mfcc & 1.22 & 0.68 & 1.76 & 0.00 \\ 
		6 & feat\_extr & db4-Mfcc & 1.62 & 1.08 & 2.16 & 0.00 \\ 
		7 & feat\_extr & db2-Mfcc & 1.77 & 1.23 & 2.31 & 0.00 \\ 
		8 & feat\_extr & db4-coif1 & 0.40 & -0.14 & 0.94 & 0.26 \\ 
		9 & feat\_extr & db2-coif1 & 0.55 & 0.01 & 1.09 & 0.04 \\ 
		10 & feat\_extr & db2-db4 & 0.15 & -0.39 & 0.69 & 0.94 \\ 
		\hline
	\end{tabular}
	\label{fig:hsd_b}
\end{table}

\subsubsection{Statistical Significance of the Difference in Predictive Performances of QDA and GB}~

\begin{table}[h]
	\centering
	\caption{HSD-test for pairwise comparison of models (Imbalanced dataset)}
	\begin{tabular}{rllrrrr}
		\hline
		& term & comparison & estimate & conf.low & conf.high & adj.p.value \\ 
		\hline
		1 & Models & KNN-LDA & 0.42 & -0.90 & 1.74 & 0.97 \\ 
		2 & Models & NB-LDA & 2.44 & 1.12 & 3.76 & 0.00 \\ 
		3 & Models & QDA-LDA & 3.96 & 2.64 & 5.28 & 0.00 \\ 
		4 & Models & RF-LDA & 5.28 & 3.96 & 6.60 & 0.00 \\ 
		5 & Models & GB-LDA & 5.46 & 4.14 & 6.78 & 0.00 \\ 
		6 & Models & SVM-LDA & 5.92 & 4.60 & 7.24 & 0.00 \\ 
		7 & Models & NB-KNN & 2.02 & 0.70 & 3.34 & 0.00 \\ 
		8 & Models & QDA-KNN & 3.54 & 2.22 & 4.86 & 0.00 \\ 
		9 & Models & RF-KNN & 4.86 & 3.54 & 6.18 & 0.00 \\ 
		10 & Models & GB-KNN & 5.04 & 3.72 & 6.36 & 0.00 \\ 
		11 & Models & SVM-KNN & 5.50 & 4.18 & 6.82 & 0.00 \\ 
		12 & Models & QDA-NB & 1.52 & 0.20 & 2.84 & 0.01 \\ 
		13 & Models & RF-NB & 2.84 & 1.52 & 4.16 & 0.00 \\ 
		14 & Models & GB-NB & 3.02 & 1.70 & 4.34 & 0.00 \\ 
		15 & Models & SVM-NB & 3.48 & 2.16 & 4.80 & 0.00 \\ 
		16 & Models & RF-QDA & 1.32 & 0.00 & 2.64 & 0.05 \\ 
		17 & Models & GB-QDA & 1.50 & 0.18 & 2.82 & 0.01 \\ 
		18 & Models & SVM-QDA & 1.96 & 0.64 & 3.28 & 0.00 \\ 
		19 & Models & GB-RF & 0.18 & -1.14 & 1.50 & 1.00 \\ 
		20 & Models & SVM-RF & 0.64 & -0.68 & 1.96 & 0.78 \\ 
		21 & Models & SVM-GB & 0.46 & -0.86 & 1.78 & 0.95 \\ 
		\hline
	\end{tabular}
	\label{fig:hsd_qda_gb_i} 
\end{table}

\clearpage
\begin{table}[h]
	\centering
	\caption{HSD-test for pairwise comparison of models (Balanced dataset)}
	\begin{tabular}{rllrrrr}
		\hline
		& term & comparison & estimate & conf.low & conf.high & adj.p.value \\ 
		\hline
		1 & Models & LDA-NB & 0.35 & -0.34 & 1.04 & 0.74 \\ 
		2 & Models & KNN-NB & 1.25 & 0.56 & 1.94 & 0.00 \\ 
		3 & Models & QDA-NB & 3.39 & 2.70 & 4.08 & 0.00 \\ 
		4 & Models & GB-NB & 3.40 & 2.71 & 4.09 & 0.00 \\ 
		5 & Models & RF-NB & 3.64 & 2.95 & 4.33 & 0.00 \\ 
		6 & Models & SVM-NB & 4.31 & 3.62 & 5.00 & 0.00 \\ 
		7 & Models & KNN-LDA & 0.90 & 0.20 & 1.59 & 0.00 \\ 
		8 & Models & QDA-LDA & 3.04 & 2.35 & 3.73 & 0.00 \\ 
		9 & Models & GB-LDA & 3.05 & 2.36 & 3.74 & 0.00 \\ 
		10 & Models & RF-LDA & 3.29 & 2.60 & 3.98 & 0.00 \\ 
		11 & Models & SVM-LDA & 3.96 & 3.27 & 4.65 & 0.00 \\ 
		12 & Models & QDA-KNN & 2.14 & 1.45 & 2.84 & 0.00 \\ 
		13 & Models & GB-KNN & 2.15 & 1.46 & 2.84 & 0.00 \\ 
		14 & Models & RF-KNN & 2.39 & 1.70 & 3.08 & 0.00 \\ 
		15 & Models & SVM-KNN & 3.06 & 2.37 & 3.76 & 0.00 \\ 
		16 & Models & GB-QDA & 0.01 & -0.68 & 0.70 & 1.00 \\ 
		17 & Models & RF-QDA & 0.25 & -0.44 & 0.94 & 0.94 \\ 
		18 & Models & SVM-QDA & 0.92 & 0.23 & 1.61 & 0.00 \\ 
		19 & Models & RF-GB & 0.24 & -0.45 & 0.93 & 0.95 \\ 
		20 & Models & SVM-GB & 0.91 & 0.22 & 1.60 & 0.00 \\ 
		21 & Models & SVM-RF & 0.67 & -0.02 & 1.36 & 0.06 \\ 
		\hline
	\end{tabular}
	\label{fig:hsd_qda_gb_b} 
\end{table}

\section{Discussion and Conclusion}~
In this paper four feature extraction techniques namely DWT-db4, DWT-db2, DWT-coif1, and MFCC were investigated and combined with seven machine learning classifiers for classifying epilepsy seizure in the case of a balanced and an imbalanced datasets. Stochastic Hold Out with 50 replications was used in the creation of the predictive performances. The two-way and one-way ANOVA test were used for statistical significance analysis of the difference in predictive performances and effect size. Tukey HSD test was used for pairwise comparison analysis of models and feature extraction methods. The results indicate that, in the imbalanced dataset, without features extraction, Gradient Boosting (GB) performed best with a classification accuracy of 95.1\% but, this accuracy is only significantly high in comparison with LDA and KNN.Among the feature extraction methods, DWT-db2 associated with Random Forest (RF) is the best combination with 98.99\% as a classification accuracy. However, this combination is challenged by DWT-db4 associated with Support Vector Machine (SVM) or Random Forest (RF) with a classification accuracy of 98.90\%.

In the balanced dataset, without features extraction, Quadratic Discriminant Analysis (QDA) performed best with a classification accuracy of 96.92\% which is only significantly high in comparison with LDA and KNN. Among the feature extraction methods, DWT-db4 associated with Support Vector Machine (SVM) is the best combination with a classification accuracy of 98.99\%. Nevertheless, this combination is challenged by DWT-db2 associated with Random Forest (RF) with a classification accuracy of 98.56\%.

The results also highlight that, MFCC performs less than all the DWT used here in the balanced or imbalanced dataset. The mean-difference are statistically significant with the minimum mean-difference of 2.01 and 1.2 when it is compared to coif1 respectively in the imbalanced and balanced dataset (Table~\ref{fig:hsd_I} and \ref{fig:hsd_b}). Whether in the balanced or the imbalanced dataset, the feature extraction methods, model, and the interaction between them have statistically significant effect on the classification accuracy. In the imbalanced dataset, the model has a medium effect while feature extraction and interaction between model and feature extraction have a large effect. In the balanced dataset, model and interaction between model and feature extraction have a large effect while feature extraction has a medium effect.

For the improvement of this work, we are planning to analyze the mean-difference of the different interactions between feature extraction methods and classifiers. Also, we are planning to study larger databases to evaluate the truthfulness of the present results in any EEG dataset of epilepsy seizure, and further, to establish the magnitude of the difference in the predictive performances between the balanced and imbalanced datasets. Another direction should be to extend the number of feature extraction methods and classifiers model, and also test our study in ECG dataset.

\printcredits


	

\bibliographystyle{cas-model2-names}
	
\bibliography{cas-refs}

\end{document}